\documentclass[aps,prb,reprint,superscriptaddress,nobibnotes]{revtex4-1}

\usepackage{amsmath} 
\usepackage{graphicx} 
\usepackage{float}
\usepackage[hidelinks]{hyperref}

\usepackage[version=4]{mhchem}

\usepackage{xcolor}
\usepackage{soul}

\usepackage{helvet}

\begin{document}

\title{
Modelling atomic and nanoscale structure in the silicon--oxygen system\\ 
through active machine learning
}

\author{Linus C. Erhard}
\affiliation{Institute of Materials Science, Technische Universit\"a{}t Darmstadt, Otto-Berndt-Strasse 3, D-64287 Darmstadt, Germany}

\author{Jochen Rohrer}
\email{rohrer@mm.tu-darmstadt.de}
\affiliation{Institute of Materials Science, Technische Universit\"a{}t Darmstadt, Otto-Berndt-Strasse 3, D-64287 Darmstadt, Germany}

\author{Karsten Albe}
\email{albe@mm.tu-darmstadt.de}
\affiliation{Institute of Materials Science, Technische Universit\"a{}t Darmstadt, Otto-Berndt-Strasse 3, D-64287 Darmstadt, Germany}

\author{Volker L. Deringer}
\email{volker.deringer@chem.ox.ac.uk}
\affiliation{Department of Chemistry, Inorganic Chemistry Laboratory, University of Oxford, Oxford OX1 3QR, United Kingdom}

\begin{abstract}
Silicon--oxygen compounds are among the most important ones in the natural sciences, occurring as building blocks in minerals and being used in semiconductors and catalysis. 
Beyond the well-known silicon dioxide, there are phases with different stoichiometric composition and nanostructured composites.
One of the key challenges in understanding the Si--O system is therefore to accurately account for its nanoscale heterogeneity beyond the length scale of individual atoms.
Here we show that a unified computational description of the full Si--O system is indeed possible, based on atomistic machine learning coupled to an active-learning workflow. 
We showcase applications to very-high-pressure silica, to surfaces and aerogels, and to the structure of amorphous silicon monoxide.
In a wider context, our work illustrates how structural complexity in functional materials beyond the atomic and few-nanometre length scales can be captured with active machine learning.
\end{abstract}

\maketitle

\section*{Introduction}
Elemental silicon and its oxide, silica (\ce{SiO2}), are widely studied building blocks of the world around us: \cite{heaneySilicaPhysicalBehavior1994a}
from minerals in geology to silicon-based computing architectures; thin-film solar cells in which amorphous silicon is the active material;\cite{yoshikawaSiliconHeterojunctionSolar2017}
or zeolite catalysts based on the \ce{SiO2} parent composition.\cite{liEmergingApplicationsZeolites2021} 
Some of these materials have a single phase and are precisely defined on the atomic scale, whereas others show longer-ranging, hierarchical structures and varying degrees of disorder. 
For example, silica aerogels contain pores with sizes of 5 to 100~nm, leading to very low thermal conductivity and making aerogels promising candidates for thermal insulation.\cite{soleimanidorchehSilicaAerogelSynthesis2008}
Under pressure, \ce{SiO2} shows amorphous--amorphous transitions to structures exceeding sixfold coordination, \cite{prescherSixfoldCoordinatedSi2017} crystallisation from the amorphous phase under shock compression, \cite{tracySituXRayDiffraction2018} and conversely the formation of complex disordered phases from crystalline \ce{SiO2}.\cite{tracyStructuralResponseAquartz2020} Beyond fundamental studies, there is much technological importance in silicon--oxygen phases with nanoscale structure---the interface between Si and \ce{SiO2} is essential in silicon metal-oxide semiconductors, and defects at this interface have been investigated for decades.\cite{termanInvestigationSurfaceStates1962, cardSiSiO2Interface1979, pantelidesSiSiO2SiC2006a}

A material in the binary silicon--oxygen system which is in fact dominated by such interfaces is the so-called silicon monoxide (SiO). The structure of SiO was controversially discussed for long; \cite{potterSiliconMonoxide1907,bradyStudyAmorphousSiO1959}
today, it is known as a nanoscopic mixture of amorphous Si and \ce{SiO2}.\cite{hohlInterfaceClustersMixture2003,schulmeisterTEMInvestigationStructure2003,hirataAtomicscaleDisproportionationAmorphous2016a} 
Initial applications of SiO have been in protective layers for mirrors \cite{hassPreparationStructureApplications1950} or dielectrics for thin-film capacitors;\cite{poatPropertiesPulsedepositedThinfilm1969} more recently, the same material has emerged as a promising anode material for lithium-ion batteries. \cite{yangSiOxbasedAnodesSecondary2002,liuSiliconOxidesPromising2019}
However, to be able to fully exploit SiO in next-generation energy-storage solutions, it would be valuable to understand the features of the nanoscopic structure on an atomistic level.

To develop atomic-scale models of complex materials such as SiO, molecular-dynamics (MD) computer simulations have become a central research tool.
While there are now plenty of interatomic potentials for silicon \cite{stillingerComputerSimulationLocal1985, tersoffNewEmpiricalApproach1988, leeModifiedEmbeddedAtom2007} and silica, \cite{vanbeestForceFieldsSilicas1990b,vashishtaInteractionPotentialSiO1990b,carreNewFittingScheme2008a} the number of potentials for the mixed ({\em i.e.}, full binary) system is limited due to its chemical complexity. \cite{yasukawaUsingExtendedTersoff1996a,vanduinReaxFFSiOReactiveForce2003,yuChargeOptimizedManybody2007,shanSecondgenerationChargeoptimizedManybody2010} 
Alongside established, empirically fitted potentials based on physical models, alternatives based on large datasets and machine learning (ML) have emerged in recent years. These models have been fitted to silicon \cite{bartokMachineLearningGeneralPurpose2018} as well as silica \cite{erhardMachinelearnedInteratomicPotential2022} and also to the more complex silica--water system.\cite{royLearningReactivePotential2023} ML potentials promise the accuracy of first-principles methods such as density-functional theory (DFT) for a small fraction of the cost. ML potentials are now firmly established in the field of computational materials science and their application to homogeneous phases has been well documented.

\begin{figure*}[t]
    \includegraphics[width=17cm]{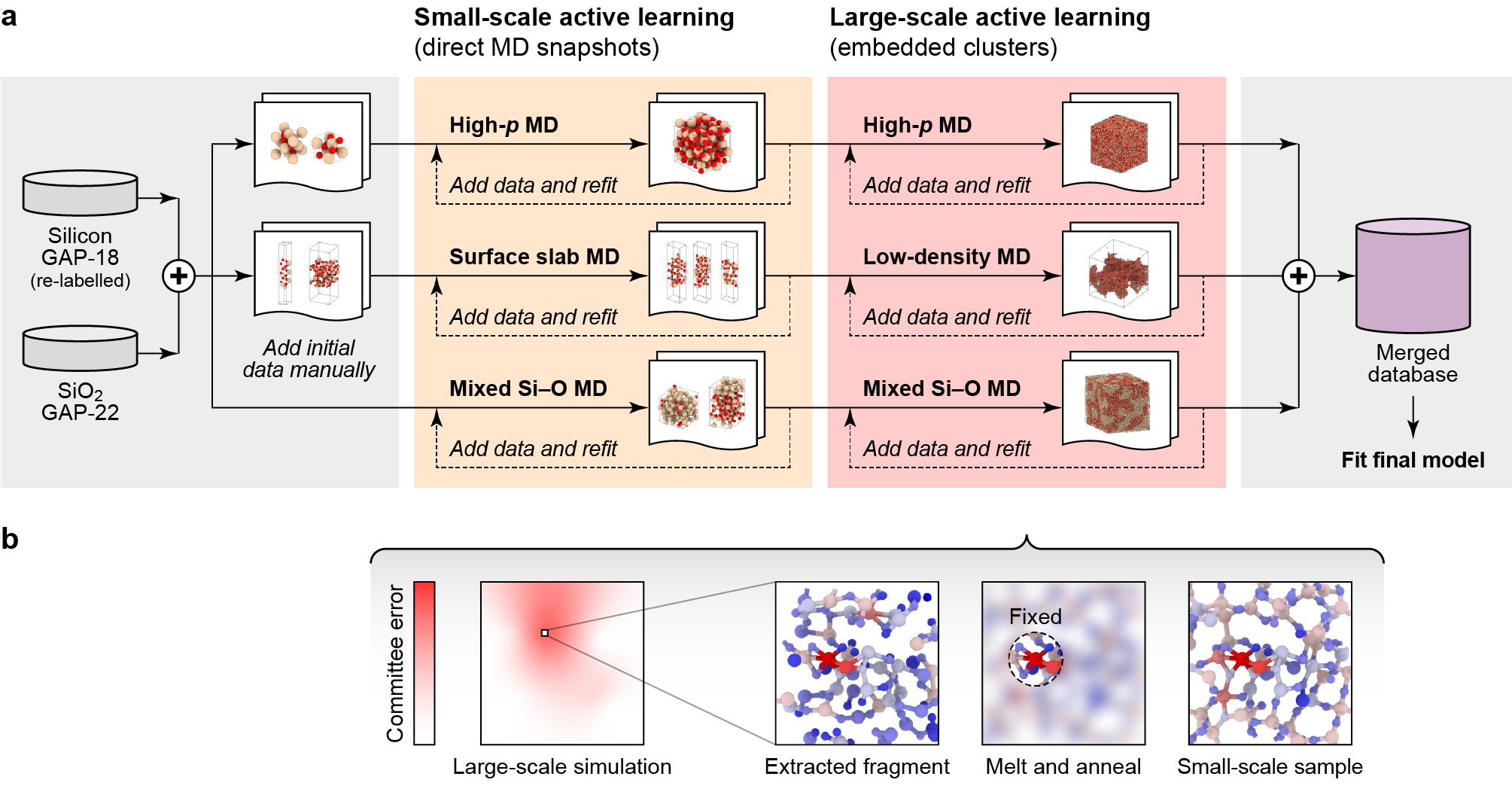}
    \caption{
    \textbf{An active-learning workflow for complex atomistic structures.}
    (\textbf{a}) Overview of the procedure to obtain the database. 
    After merging structures from the Si-GAP-18 (Ref.~\citenum{bartokMachineLearningGeneralPurpose2018}) and \ce{SiO2}-GAP-22 (Ref.~\citenum{erhardMachinelearnedInteratomicPotential2022}) databases, the process was split into three tracks, aiming to describe high--pressure silica, silica surfaces, and mixed Si--O systems with different stoichiometric compositions. 
    In each of the tracks, small-scale MD simulations were used to sample new structures by active learning.\cite{shapeevMomentTensorPotentials2016a} 
    In the last step, large-scale simulations were performed, where atoms with high uncertainty were recognised by a committee error. 
    (\textbf{b}) The concept of our amorphous matrix embedding approach. First, we extract the wider environment of an atom with high uncertainty. Then we keep the atom of interest, as well as the direct environment fixed, and we melt and anneal the outer environment. As a result, we obtain a small-scale structural model which has the atom of interest and its local environment embedded into an amorphous matrix. This sample can be fed into the training database.
    }
    \label{fig:overview}
\end{figure*}

In the present work, we describe a unified computational model for the Si--O system that we have obtained with the help of an active-learning scheme for local environments. 
We extract representative atomic environments from large-scale simulations and embed them in a melt-quenched amorphous matrix, allowing us to sample representative environments for the fitting of accurate ML potentials.
Our final model shows high accuracy across a wide configurational space including high-pressure silica, silica surfaces, and mixtures of silica and silicon.
We showcase the usefulness of the method by creating fully atomistically resolved, 10-nm-scale structure models of SiO.

\section*{Results}

\subsection*{Active learning for nanoscale structure}

We have developed a comprehensive dataset of atomistic structures and quantum-mechanical reference data for the binary Si--O system, as well as an interatomic potential fitted to that database in the atomic cluster expansion (ACE) framework. \cite{drautzAtomicClusterExpansion2019,lysogorskiyPerformantImplementationAtomic2021,bochkarevEfficientParametrizationAtomic2022} 
We initialised the protocol with two existing datasets for silicon (Bart\'ok et al., Ref.~\citenum{bartokMachineLearningGeneralPurpose2018}) and silica (Erhard et al., Ref.~\cite{erhardMachinelearnedInteratomicPotential2022}) respectively, and we then gradually explored the relevant configurational space using the active-learning workflow illustrated in Fig.~\ref{fig:overview}. 
Quantum-mechanical reference (``training'') data for energies and forces were obtained with the strongly constrained and appropriately normed (SCAN)\cite{sunStronglyConstrainedAppropriately2015} exchange--correlation functional for DFT, which shows excellent performance for elemental silicon \cite{Bonati2018} and the various silica polymorphs. \cite{erhardMachinelearnedInteratomicPotential2022} 

Our active-learning workflow follows three main tracks: high-pressure bulk silica, silica surfaces, and non-stoichiometric SiO$_x$ systems (Fig.~\ref{fig:overview}a). 
The individual tracks are kept separate during initial training, {\em i.e.}, they do not share their newly generated training data; however, in the end, all structures are merged into one comprehensive database. 

The single subtracks are further divided into stages. 
In the first stage, we added initial structures, {\em e.g.}, for crystalline high--pressure polymorphs or surfaces models. 
In the next stage, we fitted moment tensor potential (MTP) models \cite{shapeevMomentTensorPotentials2016a} to the database and used 
these MTPs to explore configurational space in MD and to identify new structures by active learning\cite{novikovMLIPPackageMoment2020}. 
Energies and forces for new structures were computed with DFT and added to the database.  
This process was iterated until the extrapolation threshold (Supplementary Material) was not exceed during the MD trajectories anymore. 

The third stage, highlighted in red in Fig.~\ref{fig:overview}a, is the most important part of our workflow, and is based on large-scale simulations in each track. 
We used 2--4 MTPs trained on the same database to estimate a per-atom committee error, as is commonly done for neural-network potentials.\cite{artrithHighdimensionalNeuralNetwork2012}
For atoms with high uncertainty (Supplementary Material), we extracted the environments into smaller, ``DFT-sized'' cells by an approach that we call amorphous matrix embedding (Fig.~\ref{fig:overview}b).
After identifying an atom with high uncertainty, we cut out a cube containing the corresponding environment of the atom. 
This cube has a size which is feasible for performing DFT computations; it is generally chosen larger than twice the cut-off of the potential.
After extracting the cube, the atoms within the cutoff radius of the atom with high uncertainty are kept fixed. 
The remaining structure is molten in an ML-MD simulation to create an amorphous matrix and smooth boundaries. 
Details of the procedure can be found in the Supplementary Material.

The final database was obtained by merging the data of all tracks together, including some additional samples such as clusters and vacancies. 
This database contains 11,428 structures with a total of $\approx$ 1.3 million atoms (Supplementary Material). 
For validation, we held out 5\% of these structures from training, selected at random.

\begin{table}[]
    \caption{\textbf{ML model performance.} We report energy root mean square error (RMSE) values in meV atom$^{-1}$ on different test sets. We characterise three ACE models, fitted to the same dataset but with increasing model complexity (Methods). `a' indicates amorphous structures. The CHIK\cite{carreNewFittingScheme2008} and GAP\cite{erhardMachinelearnedInteratomicPotential2022} generated structures are taken from Ref.~\citenum{erhardMachinelearnedInteratomicPotential2022}. }
    \centering
    \begin{tabular}{lcccc}
    \hline
    \hline
         & \ce{SiO2}-GAP & \multicolumn{3}{c}{Si--O ACE models} \\
         & (Ref.~\citenum{erhardMachinelearnedInteratomicPotential2022}) & \multicolumn{3}{c}{(This work)} \\
         \cline{3-5}
         & & Linear & F--S & Complex \\
         & & $(N=1)$ & $(N=2)$ & $(N=8)$ \\
    \hline
    \ce{SiO2} crystals & 1.0 & 0.8 & 1.1 & 0.9 \\
    \hline
    a-\ce{SiO2} (CHIK-MD) & 3.7 & 4.1 & 5.1 & 2.2 \\
    a-\ce{SiO2} (GAP-MD)  & 1.1 & 10.3 & 9.8 & 4.6 \\
    a-\ce{SiO2} (ACE-MD)  & 4.0 & 8.0 & 7.4 & 3.2 \\
    a-\ce{SiO2} surfaces & 14.9 & 21.4 & 18.0 & 4.7 \\
    \hline
    a-Si$^{a}$  & $> 1,600$ & 115.8 & 53.9 & 51.5 \\
    a-SiO$_x$$^{a}$  & $> 4,200$ & 37.8 & 35.0 & 38.0 \\
    high-$p$ a-SiO$_2$$^{a}$ & 122.7 & 15.1 & 5.6 & 4.6 \\
    \hline
    \hline
    \multicolumn{5}{p{7cm}}{\footnotesize $^{a}$Structural models generated using ACE-MD.}\\
    \end{tabular}
    \label{tab:errors}
\end{table}

\subsection*{Performance}

The final potential is a complex non-linear ACE model, obtained by summation of one linear and seven non-linear ACE terms (Methods).
This approach allows a more flexible description than just a linear or Finnis--Sinclair-like embedding, at only moderately higher computational expense.
The resulting potential has a test-set root mean square error (RMSE) of 16.7 meV atom$^{-1}$ for energies and 306 meV \AA{}$^{-1}$ for forces. 
These errors are averaged over the full dataset, however, and so they are not in themselves sufficient to characterise the quality of the potential. For example, they refer to a highly heterogeneous set of structures, with target energy values spanning more than 8 eV atom$^{-1}$, and a range of forces of 40 eV \AA{}$^{-1}$ covered by the database. 
Furthermore, the numerical accuracy of the potential in certain parts of configurational space ({\em e.g.}, crystalline polymorphs) is far more important than in others ({\em e.g.}, liquid and amorphous structures). 
In Table \ref{tab:errors}, we therefore show the performance of our model on different separate test sets.
The complex non-linear ACE is compared to our previous silica GAP model described in Ref.~\citenum{erhardMachinelearnedInteratomicPotential2022} (``\ce{SiO2}-GAP-22'' in the following), and also to simpler ACE models fitted to the new database using linear and Finnis--Sinclair-like embeddings, respectively. 
Indeed, the complex non-linear ACE potential is the only one among the three which achieves comparable errors to \ce{SiO2}-GAP-22 for amorphous and crystalline structures.
In contrast, for amorphous elemental silicon, mixed-stoichiometry as well as high-pressure phases the complex non-linear ACE is significantly more accurate than \ce{SiO2}-GAP-22, since these structures are not part of the GAP database.
This table therefore indicates the main challenge -- and its solution -- in the present model compared to the previous GAP: both are highly accurate for crystalline ($\approx 1$ meV atom$^{-1}$) and bulk amorphous ($\approx 5$ meV atom$^{-1}$) \ce{SiO2}, but our ACE model caters to a much wider range of scenarious outside of the 1:2 stoichiometric composition.

\begin{figure}
    \centering
    \includegraphics[width=8.5cm]{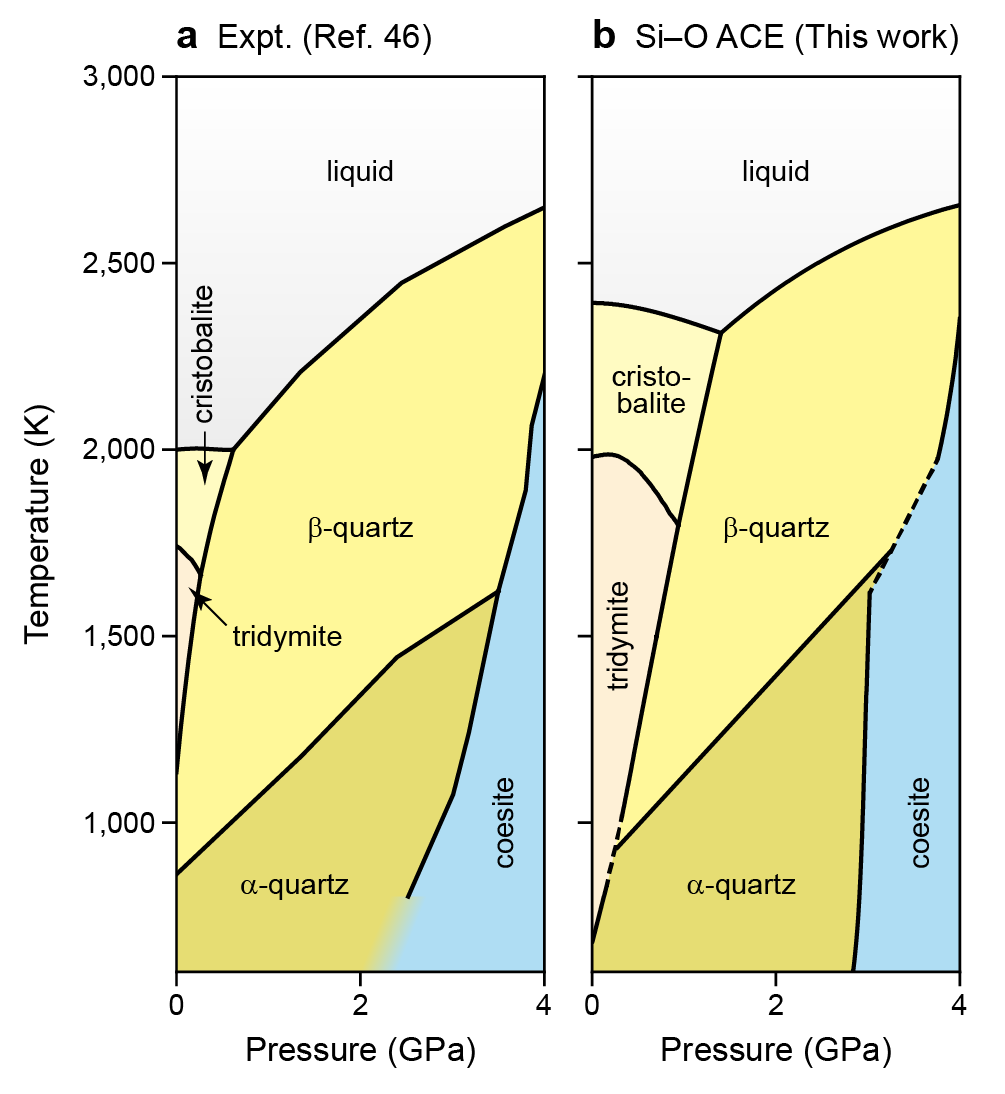}
    \caption{\textbf{Temperature--pressure phase diagram of \ce{SiO2}.} 
    (\textbf{a}) Phase diagram calculated based on experimental data, adapted from the literature (Ref.~\citenum{swamyThermodynamicAssessmentSilica1994}). 
    (\textbf{b}) The same phase diagram calculated based on predictions from our Si--O ACE  model and thermodynamic integration.}
    \label{fig:phase_diagram}
\end{figure}

\begin{figure*}
    \centering
    \includegraphics[width=0.7\textwidth]{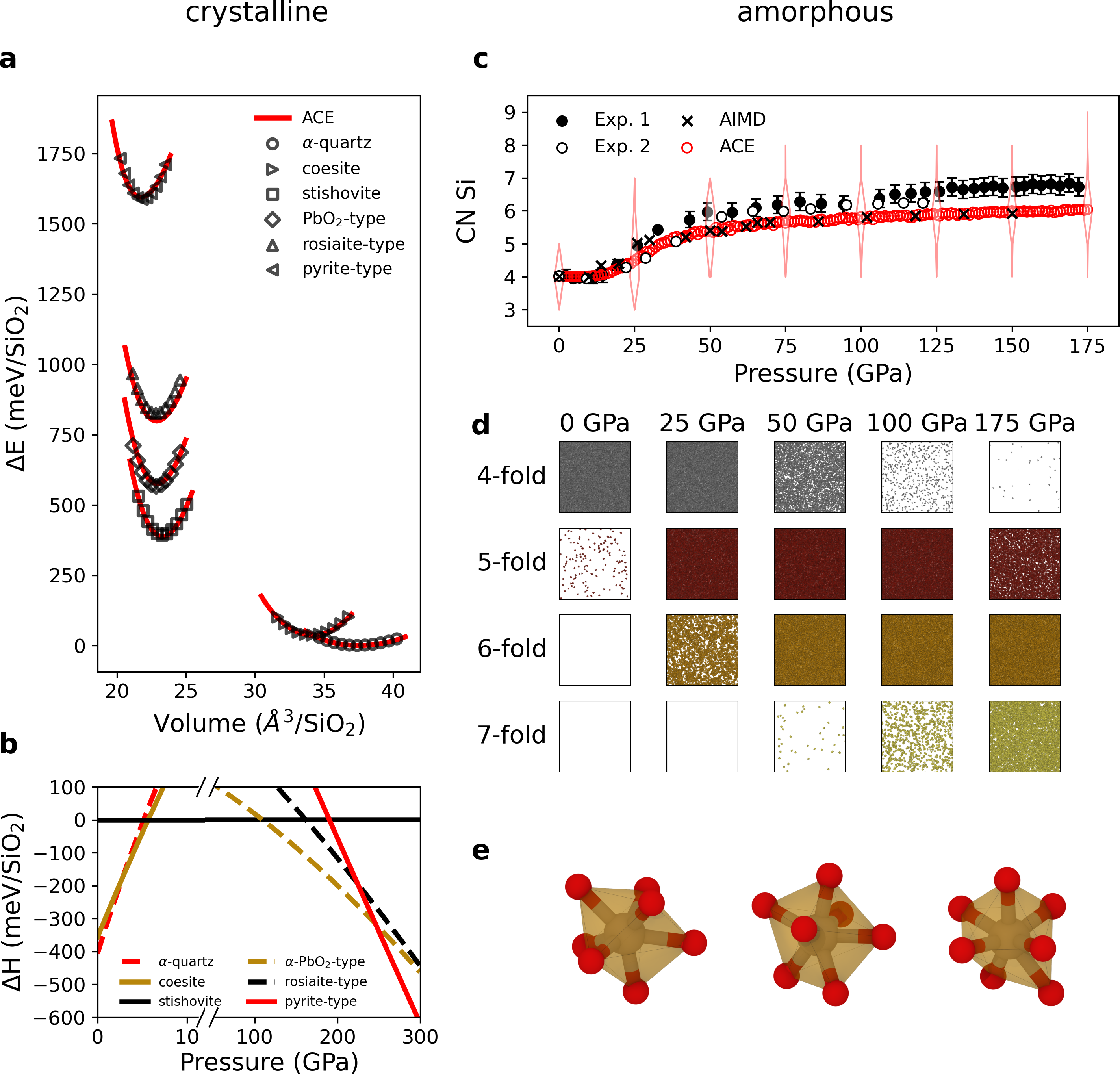}
    \caption{\textbf{Silica at megabar pressures.} (\textbf{a}) Energy--volume curves for high-$p$ silica polymorphs with energies referenced to $\alpha$-quartz. Markers indicate SCAN DFT results; lines indicate the ACE prediction. (\textbf{b}) Enthalpy differences for various silica polymorphs referenced to stishovite. (\textbf{c}) Compression of a vitreous silica structure. Results for the average silicon coordination number are compared to experimental measurements from Refs.~\citenum{prescherSixfoldCoordinatedSi2017} (''Exp. 1'') and~ \citenum{konoStructuralEvolutionMathrmSiO2020} (''Exp. 2'') and \textit{ab initio} MD simulations from Ref.~\citenum{petitgirardMagmaPropertiesDeep2019}. The distribution of coordination numbers at selected pressures is indicated by violin points. (\textbf{d}) Snapshots of the compression simulation showing coordination polyhedra for different coordination numbers (only). (\textbf{e}) Visualisation of the coordination environments of selected 7-fold coordinated silicon atoms.}
    \label{fig:high_pressure}
\end{figure*}

\subsection*{Phase diagram of SiO$_2$}

Figure \ref{fig:phase_diagram} shows the phase diagram of \ce{SiO2} calculated by thermodynamic integration\cite{menonAutomatedFreeenergyCalculation2021,dekoningOptimizedFreeEnergyEvaluation1999} using the ACE potential compared to a CALPHAD phase diagram from literature.\cite{swamyThermodynamicAssessmentSilica1994}
The ACE and CALPHAD predictions agree well throughout, and 
for the boundary between quartz and coesite we observe almost quantitative agreement. 
In contrast, the cristobalite and tridymite phases seem to be over-stabilised. 
At 0~GPa, the melting point is notably overestimated (about 2,400 K, compared to $\approx 2,000$ K experimentally \cite{swamyThermodynamicAssessmentSilica1994}); moreover, the phase stability regions of both phases are more extended than in the reference.
To illustrate the sensitivity of the analysis to small errors in predicted energies, we added a fictitious energy penalty of 5 meV atom$^{-1}$ for cristobalite and tridymite (Supplementary Fig.~S1a);
in this case, the transition lines already agree much better with the CALPHAD reference than before. 
Further numerical tests showed that the tridymite--cristobalite transition line, in particular, is strongly affected by small shifts in energy (Supplementary Fig.~S1b--f).
We thus conclude that the quantitative deviation seen in Fig.~\ref{fig:phase_diagram} is due to the inaccuracy of the underlying exchange--correlation functional, rather than indicating a shortcoming of the ACE approach.

\subsection*{High--pressure structural transitions of SiO$_2$}

Figure \ref{fig:high_pressure} characterises high-pressure properties of silica.
In Fig.~\ref{fig:high_pressure}a, we show energy--volume curves of $\alpha$-quartz, coesite, stishovite, $\alpha$-PbO$_2$-type, and pyrite-type silica as predicted by our ACE model and compared with DFT data, with which they agree well.
In addition, we tested the behaviour of the model for rosiaite-type silica, which was recently observed in experiment \cite{otzenEvidenceRosiaitestructuredHighpressure2023} and predicted theoretically \cite{tsuchiyaNewHighpressureStructure2022} for direct compression of $\alpha$-quartz. 
In contrast to the structures mentioned before, this particular polymorph is not part of the training database. 
Nevertheless, the ACE model reproduces DFT data for this structure similarly well as for the other polymorphs.

Figure \ref{fig:high_pressure}b shows an enthalpy--pressure diagram at 0~K. 
For lower pressures, there is a transition from $\alpha$-quartz to coesite between 2.5 and 3.0~GPa, consistent with the predicted phase diagram (Fig.~\ref{fig:phase_diagram}),
followed by a transition to stishovite at 5.5-6.0~GPa. 
At higher pressures of $\approx 110$ GPa, we observe the transition from stishovite to $\alpha$-PbO$_2$-type silica. 
Experimentally, rather than stishovite (rutile type), the structurally closely related \ce{CaCl2} (distorted rutile) type polymorph of silica is stable. 
The transition from CaCl$_2$- to $\alpha$-PbO$_2$-type silica was observed at 120~GPa and 2400~K.\cite{murakamiStabilityCaCl2typeAPbO2type2003}
Given that our enthalpy data correspond to a temperature of 0~K, both values agree well with each other.
For the transition of $\alpha$-PbO$_2$- to pyrite-type silica, our ACE model predicts a pressure of $\approx 246$~GPa, in good agreement with the experimentally determined transition pressure of $\approx 260$~GPa at 1800~K.\cite{kuwayamaPyriteTypeHighPressureForm2005} 
Finally, rosiaite-type silica \cite{otzenEvidenceRosiaitestructuredHighpressure2023} is correctly identified as metastable over the pressure range studied.

Figure \ref{fig:high_pressure}c shows the pressure evolution of the average coordination number (CN) of silicon in amorphous silica, extracted from an MD simulation at room temperature and under isostatic pressure.
The ACE results agree well with experiment up to about 50~GPa,\cite{konoStructuralEvolutionMathrmSiO2020,prescherSixfoldCoordinatedSi2017} and with \textit{ab initio} MD\cite{petitgirardMagmaPropertiesDeep2019} results over the whole pressure range.
The good agreement with experiment is particularly true for the data from Ref. \citenum{konoStructuralEvolutionMathrmSiO2020}. 
Above 50~GPa, the ACE underestimates the average CN: 
at 175~GPa the experiment predicts it to be about 7; the ACE simulation predicts it to be 6.
Importantly, this does not mean that there are no sevenfold-coordinated environments, but there is a residual number of 5-fold coordinated atoms as well, lowering the average (Fig.~\ref{fig:high_pressure}d).
A possible reason for the good agreement with the \textit{ab initio} result, but the deviation from the experimental results, might be the limited time scales in our simulations, which hinder a complete transition into higher-coordinated environments. 
Moreover, we note that calculated X-ray-Raman spectra of the \textit{ab initio} structures from Ref.~\citenum{petitgirardMagmaPropertiesDeep2019} are in good agreement with experiment indicating a lower CN.
Other MD simulations also showed slightly lower CNs than the experimental values. \cite{murakamiUltrahighpressureFormSi2019}
Figure \ref{fig:high_pressure}e shows three different 7-fold coordinated environments extracted from the simulations.
A CN of 7 in amorphous silica might be surprising since silicon is sixfold-coordinated in all crystalline silica polymorphs that are stable in this pressure range. 
However, the pyrite-type phase, which becomes thermodynamically stable at $\approx$ 240--260~GPa contains silicon atoms with a 6+2-fold environment. 
A recent study found certain, but limited, similarities between these seven-fold environments in the glass and pyrite silica.\cite{murakamiUltrahighpressureFormSi2019}

In Supplementary Fig.~S2, we show two additional structural fingerprints which have been commonly analysed in experiment: the position of the first sharp diffraction peak and the Si--O bond length. 
For both cases, our simulations show good agreement with experiment.

\begin{figure}
    \centering
    \includegraphics[width=85mm]{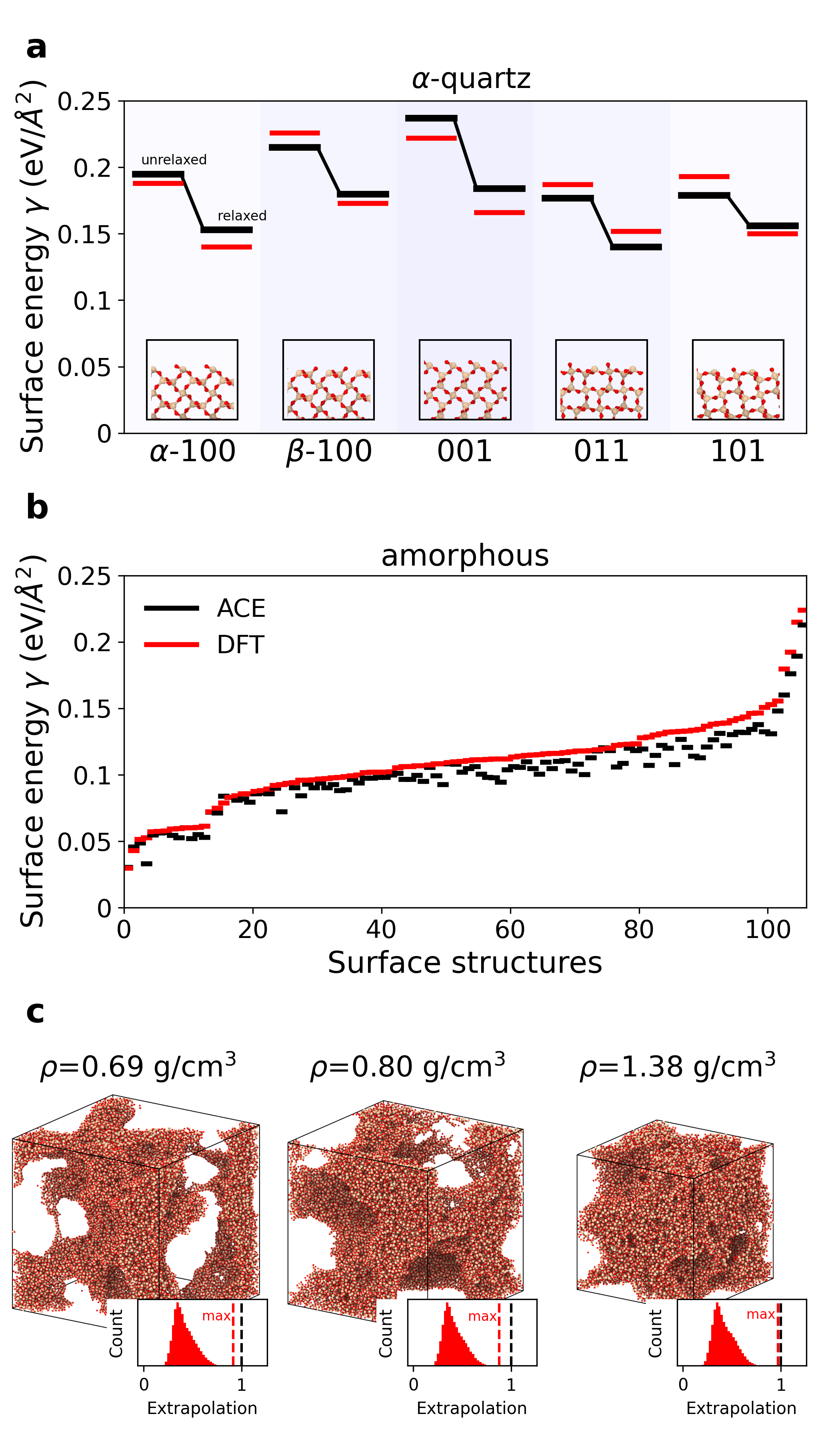}
    \caption{
    \textbf{Surfaces and aerogels.}
    (a) Surface energies of various $\alpha$-quartz surfaces in relaxed and unrelaxed state. The black lines are the ACE results while the red lines are the DFT single point results for the relaxed/unrelaxed structures. The structure pictures are showing the unrelaxed surfaces. (b)~Surface energies of amorphous models calculated with ACE and DFT. Amorphous structures were taken from Ref. \citenum{erhardMachinelearnedInteratomicPotential2022}, and have been created by various interatomic potentials: the BKS,\cite{vanbeestForceFieldsSilicas1990a} CHIK,\cite{carreNewFittingScheme2008} \ce{SiO2}-GAP-22,\cite{erhardMachinelearnedInteratomicPotential2022} Munetoh,\cite{munetohInteratomicPotentialSi2007b} and Vashishta\cite{vashishtaInteractionPotentialSiO1990a} models. The structures were relaxed with ACE and sliced at various points, followed by another ACE relaxation. DFT surface energies were determined without further relaxation. (c)~Exemplary porous amorphous silica structures with various densities. Additionally, we show the distribution of the linear extrapolation grade\cite{lysogorskiyActiveLearningStrategies2023} and the maximum extrapolation grade (red line) for each structure. For all structures, the maximum extrapolation grade of the atoms is below one. 
    }
    \label{fig:surfaces}
\end{figure}

\subsection*{SiO$_2$ surfaces and aerogels}

Figure \ref{fig:surfaces} tests the ability of the potential to accurately predict surface energies. 
We begin with validation for different $\alpha$-quartz surfaces: we created surface slab models, relaxed them with the ACE models, and evaluated the energetics, and therefore the surface energy per area, with DFT single-point computations (Fig.~\ref{fig:surfaces}a).
The ACE results agree well with DFT, especially considering that the training database does not contain all the surface terminations shown.
Whilst these surface energies {\em can} be computed with DFT, realistic amorphous surface energies are much harder to calculate due to the required system sizes. 
Therefore, Fig. \ref{fig:surfaces}b validates the potential on 125 small--scale surface structures of amorphous \ce{SiO2}, each containing 192 atoms. The surface models are created based on bulk structures from Ref.~\citenum{erhardMachinelearnedInteratomicPotential2022};
the latter had been generated in melt--quench simulations with different interatomic potentials and therefore span a range of energies.
Regardless of the starting structure, the ACE model captures the surface energy for all surfaces models very well: the total RMSE is about 0.01 eV/\AA{}$^{2}$, and only a slight underestimation compared to DFT is seen.
Moreover, there are no clear outliers although the various surface energies indicate a large diversity of the surface structures.

\begin{figure*}
    \centering
    \includegraphics[width=17cm]{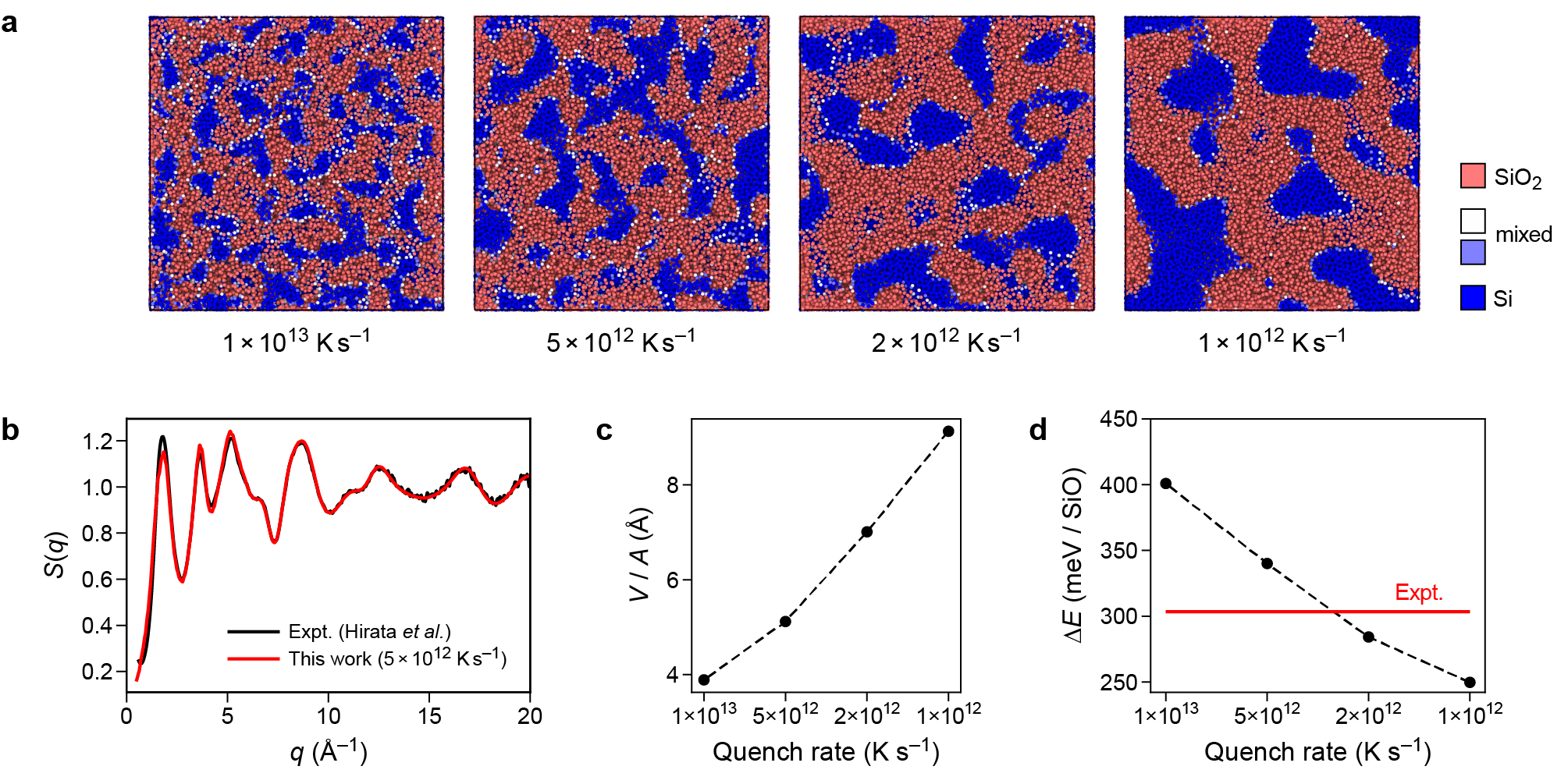}
    \caption{
    \textbf{Nanoscale segregation in amorphous silicon monoxide.}
    (\textbf{a}) Visualisation of SiO structures generated by quenching from the melt at rates between $10^{13}$ and $10^{12}$ K s$^{-1}$.
    Colour-coding is based on the nearest-neighbour Si--O coordination numbers, which are four in \ce{SiO2} and zero in elemental silicon. Accordingly, \ce{SiO2}-like and Si-like regions are indicated in red and blue, respectively.
    (\textbf{b}) Structure factor for the $5 \times 10^{12}$ K s$^{-1}$ simulation.
    (\textbf{c}) Relation between grain volume of the silicon grains and the interface area between silicon and silica. With increasing quench rate, the grain size of the structures decreases. 
    (\textbf{d}) Energy of the SiO structures referenced to $\alpha$-quartz and to diamond-type silicon, compared to the experimental standard enthalpy of formation for SiO. \cite{nagamoriGibbsFreeEnergies1995}}
    \label{fig:sio}
\end{figure*}

The amorphous surfaces shown are already very complex, but in reality they are often not flat as here.
They have curvature, for example when occurring inside pores, and such complex structures can no longer be directly validated with DFT. 
Fig.~\ref{fig:surfaces}c therefore shows how well atomic environments in various porous amorphous structures are covered by the dataset. 
These structures were prepared by straining amorphous structures at elevated temperatures to the desired density.
To validate the performance of the potential on this model, we show the linear extrapolation grade according to the maxvol selection.\cite{podryabinkinActiveLearningLinearly2017,lysogorskiyActiveLearningStrategies2023}
An extrapolation grade above 1 corresponds to atomic environments that have not been covered in the training database.
This does not mean that the potential is no longer reliable, as there is a certain range of more or less reliable extrapolation, but as the extrapolation grade increases, non-physical behaviour and failure of the potential becomes more and more likely.\cite{novikovMLIPPackageMoment2020} 
For all porous structures, regardless of density, we find that the maximum extrapolation value is less than 1.
Thus, we observe no extrapolation in any of the considered cases.
This indicates an accurate description of the potential for a variety of curved surfaces.

\subsection*{Elemental silicon}

Whilst our ACE model is designed for the binary Si--O system, we show in Table \ref{tab:ace_scan_silicon} the performance for diamond-like elemental silicon compared to both DFT and experiment.
The bulk modulus of diamond is very well reproduced, whereas the 
vacancy formation energy is underestimated by about 30~\%. 
The experimental surface energies are well recovered by the ACE model, but this may be partly due to serendipity, because the SCAN ground-truth data show poorer quality (Table \ref{tab:ace_scan_silicon}).
We also computed the linear thermal expansion coefficient of diamond-type silicon in the quasiharmonic approximation. We found almost perfect agreement between the ACE prediction and experiment; in particular, the unusual negative expansion coefficient below 130 K is reproduced (Supplementary Fig.~S3). 
Finally, melt--quench simulations were performed to generate amorphous silicon structures (Supplementary Fig.~S4).
The agreement with the experimental structure factor is as good as that for a GAP-18-generated structure from Ref.~\citenum{deringerRealisticAtomisticStructure2018a}.
In addition, we are able to achieve lower quenching rates with the ACE than with the GAP, and for quenching rates as low as 10$^{10}$ K/s, we observed crystallisation. 

Compared to Si-GAP-18,\cite{bartokMachineLearningGeneralPurpose2018} we observe higher errors with respect to the reference data (Table \ref{tab:ace_scan_silicon}). 
This is not surprising as our database contains two elements and has a strong focus on the configurational space of SiO$_2$. 
However, for properties such as the bulk modulus, the agreement with experiment is better than that of GAP-18.
The reason for this is in the underlying SCAN data, which appears to provide a more accurate description of silicon than the GGA functionals used previously. 
Due to the lower accuracy in reproducing the SCAN data, the potential has some shortcomings for higher pressure polymorphs: the bc8 phase is erroneously predicted to be stable at elevated pressure (Supplementary Fig.~S5).
However, very-high-pressure silicon phases were not the scope of the present work -- instead, the focus in this case is on the accurate description of ambient-pressure silicon environments as a constituent part of mixed binary phases and nanostructures.

\begin{table}
    \centering
    \caption{\textbf{Properties of diamond-type silicon.} We show SCAN DFT values for reference, as well those obtained with the complex ACE potential; both computations are compared to the GAP-18 model, the corresponding reference data (PW91) and to experimental data (``Expt.'').}
    \begin{tabular}{lccccccc}
    \hline
    \hline
         & & \multicolumn{2}{c}{This work} & \multicolumn{2}{c}{Ref.~\citenum{bartokMachineLearningGeneralPurpose2018}} & \\
         \cline{3-4} \cline{5-6}
         Property & & SCAN & ACE & PW91 & GAP & Expt.  \\
         \hline
         Bulk modulus & (GPa) & 100.0 & 100.8 & 88.8 & 88.4 & 97.8 \cite{hallElectronicEffectsElastic1967} \\
         \hline
         Vacancy $E_f$ & (eV) & 4.09 & 2.80 & 3.67 & 3.61 & 4 \cite{fukataVacancyFormationEnergy2001} \\
         \hline
         $\gamma_{100}$ & (eV/\AA$^2$) & 0.155 & 0.117 & 0.135 & 0.133 & 0.133 \cite{jaccodineSurfaceEnergyGermanium1963} \\
         $\gamma_{110}$ & (eV/\AA$^2$) & 0.126 & 0.090 & 0.095 & 0.094 & 0.094 \cite{jaccodineSurfaceEnergyGermanium1963} \\
         $\gamma_{111}$ & (eV/\AA$^2$) & 0.113 & 0.076 & 0.098 & 0.096 & 0.077 \cite{jaccodineSurfaceEnergyGermanium1963} \\
         
    \hline
    \hline
    \end{tabular}
    \label{tab:ace_scan_silicon}
\end{table}

\subsection*{SiO and mixed silicon--silica systems}

Whilst the results so far have served to demonstrate the usefulness of the appraoch -- both in terms of development of efficiently generated datasets and the fitting within the ACE framework -- we are now in a position to study an actual application problem.
To this end, Fig.\ \ref{fig:sio}a shows structural models of SiO.
Experimentally, amorphous SiO is obtained by deposition of SiO from the gas phase.\cite{fergusonVaporPressureSilicon2008}
In contrast, we created our models by melt--quench simulations.
SiO phases are known to be metastable with respect to Si and SiO$_2$. 
For example, a recent DFT-based crystal-structure prediction study explored possible ordered phases of homogeneous SiO, and found for a range of structures that these are metastable compared to the crystalline mixture of Si and SiO$_2$.\cite{alkaabiSiliconMonoxideAtm2014} 
We verified that our ACE potential similarly reproduces the metastability of the ambient pressure phases (Supplementary Fig.~S6).
In good agreement with these results, our melt-quenched structures show a clear segregation between amorphous silicon (blue) and amorphous silica (red).
With decreasing quench rate, the number of silicon grains decreases while the grain size increases. 
Figure \ref{fig:sio}b shows the structure factor, $S(Q)$, determined at 300~K for the structure quenched with $5\times 10^{12}$~K/s.
The $S(Q)$ data for the other structures are shown in Supplementary Fig.~S7.
The structure factor of the structures generated by a quenching rate of $5\times 10^{12}$~K/s agrees best with the experimental structure factor from Ref. \citenum{hirataAtomicscaleDisproportionationAmorphous2016}.
Figure \ref{fig:sio}c shows the ratio between the volume of the silicon grains divided by the interface area; details are given in the Methods section.
In the approximation of spherical particles, the grain diameter is $d=6\cdot V_{\text{Si,grains}}/A_{\text{interface}}$.
From this we can estimate average grain diameters between 24~\AA~and 54~\AA~for the tested quench rates.
These grain diameters agree very well with transmission electron microscopy measurements, which indicated diameters of 30 to 40~\AA.\cite{schulmeisterTEMInvestigationStructure2003}

Figure \ref{fig:sio}d shows the excess energies of the structures referenced to to $\alpha$-quartz and diamond-type silicon. 
The SiO structures were relaxed by optimisation of the cell size as well as the atom positions at 0~K. 
As experimental reference, we show the standard enthalpy of formation of SiO.\cite{nagamoriGibbsFreeEnergies1995}
The structures generated by quench rates of 5$\times$10$^{12}$ and 2$\times$10$^{12}$ K/s have energies comparable to experiment. 
Indeed, we can even create structures that are energetically {\em more} favourable than in experiment, noting again that our procedure to produce the structures deviates significantly from the experimental one.

But is this really an improvement compared to existing, empirically fitted interatomic potentials? 
Indeed, there are already two potentials implemented in \texttt{LAMMPS}\cite{thompsonLAMMPSFlexibleSimulation2022} which are able to perform simulations of mixed systems: the Munetoh potential \cite{munetohInteratomicPotentialSi2007b} and a charge-optimised many-body (COMB) potential. \cite{shanSecondgenerationChargeoptimizedManybody2010}
For both potentials, we tested the same procedure to generate structural models of SiO.
The Munetoh potential yielded a homogeneous structure without observable segregation into silicon and \ce{SiO2},
and the resulting structure factor (Supplementary Fig.~S7) deviates strongly from experiment. 
For the COMB potential, we observed pore formation at elevated temperatures, finally resulting in a strongly increased simulation-cell size. 
Therefore, we only equilibrated our best-matching structure at room temperature and analysed the change in structure factor (Supplementary Fig.~S6): 
again, we observed a strong deviation from experiment, indicating that the structure is very different from the ACE model prediction.

\section*{Discussion} 
Understanding the microscopic nature of interfaces and nanostructured matter is essential to advancing materials research. 
Here, we have presented an active-learning scheme that we term ``amorphous matrix embedding'' that can realistically represent environments from large-scale simulations in DFT-accessible cells, enabling fast and accurate atomistic modelling of heterogeneous materials.
We used the approach to develop a general-purpose interatomic potential for binary Si--O phases with varied compositions that is able to describe the trifecta of modelling challenges in this material system: very high pressure phases (relevant to geology), surfaces (relevant to catalysis), and mixed stoichiometric compositions with nanoscale heterogeneity (relevant to battery systems).

Using the ACE approach, we observe a speed-up of about two orders of magnitude compared to the more established GAP framework. 
This makes it possible to access long time scales and large length scales with DFT-like accuracy.
Of course, there are some shortcomings of this potential, {\em e.g.}, the lower accuracy for pure silicon compared to the state of the art -- but this use-case is not the focus of our work, as there are already competitive GAP and ACE models available. \cite{bartokMachineLearningGeneralPurpose2018,lysogorskiyPerformantImplementationAtomic2021}
In our case, they quality of the underlying meta-GGA data might cause an out-performance compared to earlier ML-potentials fitted with more economical GGA labels.

We hope that our work, and the dataset and resources developed therein, will advance the modelling of porous silica nanostructures as well as of high-pressure silica. 
For the Si--\ce{SiO2} interface, alternative interatomic potential models are scarce and the higher quality potentials come with an expensive charge-equilibration term. 
Our tests showed that the ACE describes silicon monoxide in much closer agreement with experiment than existing empirical models.

We view the present database and ML potential model as a starting point for wider-ranging studies in this important material system.
In the future, higher accuracy for the mixed system might be achieved by using charge-equilibration schemes coupled with ML potentials.\cite{koFourthgenerationHighdimensionalNeural2021}
However, this would come with much longer computing times as well as worse scaling for larger systems. 
Moreover, in the future, we will include lithium in the potential to investigate the battery performance of SiO on the atomistic scale.

\section*{Methods}

\small

\textbf{Machine-learning potential fitting.} 
We used two frameworks for fitting ML potential models. 
While constructing the reference database, we used Moment Tensor Potentials \cite{shapeevMomentTensorPotentials2016a} with active learning \cite{podryabinkinActiveLearningLinearly2017} as implemented in the \texttt{MLIP} package. \cite{novikovMLIPPackageMoment2020} 
For the final potential fit, we used the nonlinear Atomic Cluster Expansion (ACE) \cite{drautzAtomicClusterExpansion2019,lysogorskiyPerformantImplementationAtomic2021} as implemented in \texttt{PACEMAKER}. \cite{bochkarevEfficientParametrizationAtomic2022} 
For ACE, we tested a range of combinations of embeddings, and found the following to be suitable:
\begin{align*}
  E_i = \phi + \sqrt{\phi} + \sum_{i} \phi^{f_i},   
\end{align*}
with $\phi_i$ being atomic properties, which are expanded by the ACE basis functions (for details see Ref. \cite{drautzAtomicClusterExpansion2019}). 
The exponents of the embeddings include fractional exponents and higher integer powers of $f_i \in \left\{ 1/8, 1/4, 3/8, 3/4, 7/8, 2 \right\}$.
We found that especially fractions between 0 and 1 improved the behaviour of the potential.
This approach goes beyond the previously suggested linear embedding (only the first term) and Finnis--Sinclair (the first two terms) type embedding, \cite{lysogorskiyPerformantImplementationAtomic2021}
and is referred to as ``complex'' embedding in Table \ref{tab:errors}.
For the expansion of the atomic properties $\phi_i$ we used 600 basis functions with 5700 parameters.
As radial basis we employed Bessel functions. 
A $\kappa$ value of 0.01, which gives the ratio between force and energy weights value, was used during fitting.
For optimisation we used the BFGS algorithm for 2000 steps.

\textbf{DFT computations.} 
All DFT computations were performed using VASP \cite{kresseEfficiencyAbinitioTotal1996,kresseEfficientIterativeSchemes1996} 
and the projector augmented-wave method. \cite{blochlProjectorAugmentedwaveMethod1994,kresseUltrasoftPseudopotentialsProjector1999} For calculations we used the SCAN functional \cite{sunStronglyConstrainedAppropriately2015} with an energy cutoff of 900 eV and a $k$-spacing of 0.23 \AA$^{-1}$. Surface calculations were performed with dipole corrections. We note that these convergence parameters are optimised for silica; however, we found them to be also well converged for mixed phases and for pure silicon structures. Only for very-high-pressure silicon allotropes, a higher $k$-spacing would provide a relevant advantage; however, since these are not in the scope of the present work, we neglect these inaccuracies.

\textbf{MD workflows.} 
Simulation protocols were implemented using the atomic simulation environment (ASE) \cite{larsenAtomicSimulationEnvironment2017} and the OVITO Python interface. \cite{stukowskiVisualizationAnalysisAtomistic2010} 
While optimisation and small-cell MD were partially performed with ASE, large-scale MD and statics simulations were carried out using LAMMPS. \cite{thompsonLAMMPSFlexibleSimulation2022}
The time step was 1~fs.
For NVT simulations, we used a Nos\'e--Hoover thermostat with temperature damping constant of 100~fs; for NPT simulations, we added a Nos\'e--Hoover barostat with a pressure damping constant of 1,000~fs. 

Amorphous structural models as starting point for compression simulations were created by the melt-quench procedure described in Ref.~\citenum{erhardMachinelearnedInteratomicPotential2022}, now using the ACE potential.
The compression was performed under isostatic conditions. 
In each step, the pressure was initially increased by 1~GPa within 2.5~ps of simulation time,
followed by equlibration over 2.5~ps at the new pressure.
This procedure was iteratively repeated. 
Coordination numbers were determined after equilibration.

The aerogel structures were created by a similar protocol as in Ref.~\citenum{erhardMachinelearnedInteratomicPotential2022}.
An initial structure was randomized at 6,000~K for 10~ps,
instantly cooled to 4,000~K and kept there for 100~ps.
From this temperature, the liquid was cooled to 300~K with a quench rate of $10^{13}$~K s$^{-1}$. 
During the equilibration at 4,000~K and up to half of the quenching process, the cells were additionally extended to the desired density.

The mixed structures were created using the same protocol as in Ref.~\citenum{erhardMachinelearnedInteratomicPotential2022} for producing amorphous structures. 
The volume of the silicon grains was determined within OVITO by deleting all silicon atoms and usage of the \texttt{ConstructSurfaceMesh} modifier on the remaining oxygen atoms. 
The corresponding interface area was extracted in the same way.

\textbf{Phase diagram calculations.} 
Thermodynamic integration was carried out as implemented in \texttt{calphy}. \cite{menonAutomatedFreeenergyCalculation2021,dekoningOptimizedFreeEnergyEvaluation1999}
We used 50,000 equilibration steps, 800,000 switching steps for the switching to the Einstein crystal, as well as 300 steps/K for the thermodynamic integration to calculate the temperature dependence. 
Due to numerical issues, we fixed the spring constants of the Einstein crystal to 2~eV/\AA$^2$ for oxygen and 4~eV/\AA$^2$ for silicon. 
We carefully checked the influence of this constraint on the final results and found it to be negligible.

\textbf{Structure factors.}
Faber--Ziman structure factors were obtained by summation of the Fourier transformations of the partial radial distribution functions calculated with OVITO. 
The corresponding partial structure factors were weighted by atomic form factors taken from Ref.~\citenum{princeInternationalTablesCrystallography2004}. 
For the high-pressure structures, we used a cut-off radius of 20~\AA~for the radial distribution function, and analysed a single snapshot (without time averaging).
For the SiO structures, we used a cut-off of 80~\AA~and an average over 10 snapshots.

\normalsize

\section*{Data availability}

The potential parameter files, the reference data with SCAN labels, and additional supporting data will be provided openly through Zenodo upon journal publication.

\section*{Acknowledgements}
L.C.E. thanks Niklas Leimeroth for useful discussions. L.C.E. acknowledges support from the German Academic Exchange Service (Forschungsstipendien f\"ur Doktorandinnen und Doktoranden) and the Erasmus+ programme for support of two research stays at the University of Oxford.
The research was supported by the Bundesministerium für Bildung und Forschung
(BMBF) within the project FESTBATT under Grant No. 03XP0174A. 
J.R. and K.A. acknowledge support by the Deutsche Forschungsgemeinschaft (DFG, Grant no. RO 4542/4-1 and STU 611/5-1). L.C.E. acknowledges helpful discussion within the DFG GRK-2561 MatCom-ComMat.
The authors gratefully acknowledge the computing time provided to them at the NHR Center NHR4CES at TU Darmstadt (project number 01539 and p0020142). This is funded by the Federal Ministry of Education and Research, and the state governments participating on the basis of the resolutions of the GWK for national high performance computing at universities (www.nhr-verein.de/unsere-partner).

\section*{Author Contributions}
L.C.E. performed all computations and analysis, with guidance from J.R., K.A., and V.L.D. All authors contributed substantially to the design of the research and to the interpretation of the results. L.C.E. and V.L.D. wrote the paper with input from all authors.

\section*{Competing interests}

The authors declare no competing interests.

\bibliographystyle{naturemag}

\bibliography{SiO_main}

\end{document}


\title{
Supplementary Material for \\[2mm]
``Modelling atomic and nanoscale structure in the silicon--oxygen system\\ through active machine learning''
}

\author{Linus C. Erhard}
\affiliation{Institute of Materials Science, Technische Universit\"a{}t Darmstadt, Otto-Berndt-Strasse 3, D-64287 Darmstadt, Germany}

\author{Jochen Rohrer}
\email{rohrer@mm.tu-darmstadt.de}
\affiliation{Institute of Materials Science, Technische Universit\"a{}t Darmstadt, Otto-Berndt-Strasse 3, D-64287 Darmstadt, Germany}

\author{Karsten Albe}
\email{albe@mm.tu-darmstadt.de}
\affiliation{Institute of Materials Science, Technische Universit\"a{}t Darmstadt, Otto-Berndt-Strasse 3, D-64287 Darmstadt, Germany}

\author{Volker L. Deringer}
\email{volker.deringer@chem.ox.ac.uk}
\affiliation{Department of Chemistry, Inorganic Chemistry Laboratory, University of Oxford, Oxford OX1 3QR, United Kingdom}

\maketitle
\renewcommand{\figurename}{Supplementary Figure}
\renewcommand{\tablename}{Supplementary Table}
\renewcommand{\thetable}{\arabic{table}}
\let\oldthefigure\thefigure
\renewcommand{\thefigure}{S\oldthefigure}
\let\oldthetable\thetable
\renewcommand{\thetable}{S\oldthetable}

\onecolumngrid
The Supplementary Material is split into two parts. In the first part we present figures referenced in the main article. In the second part methodological details about the database generation, the active learning and calculation of various properties are given. 
\twocolumngrid

\begin{figure*}
    \centering
    \includegraphics[width=17cm]{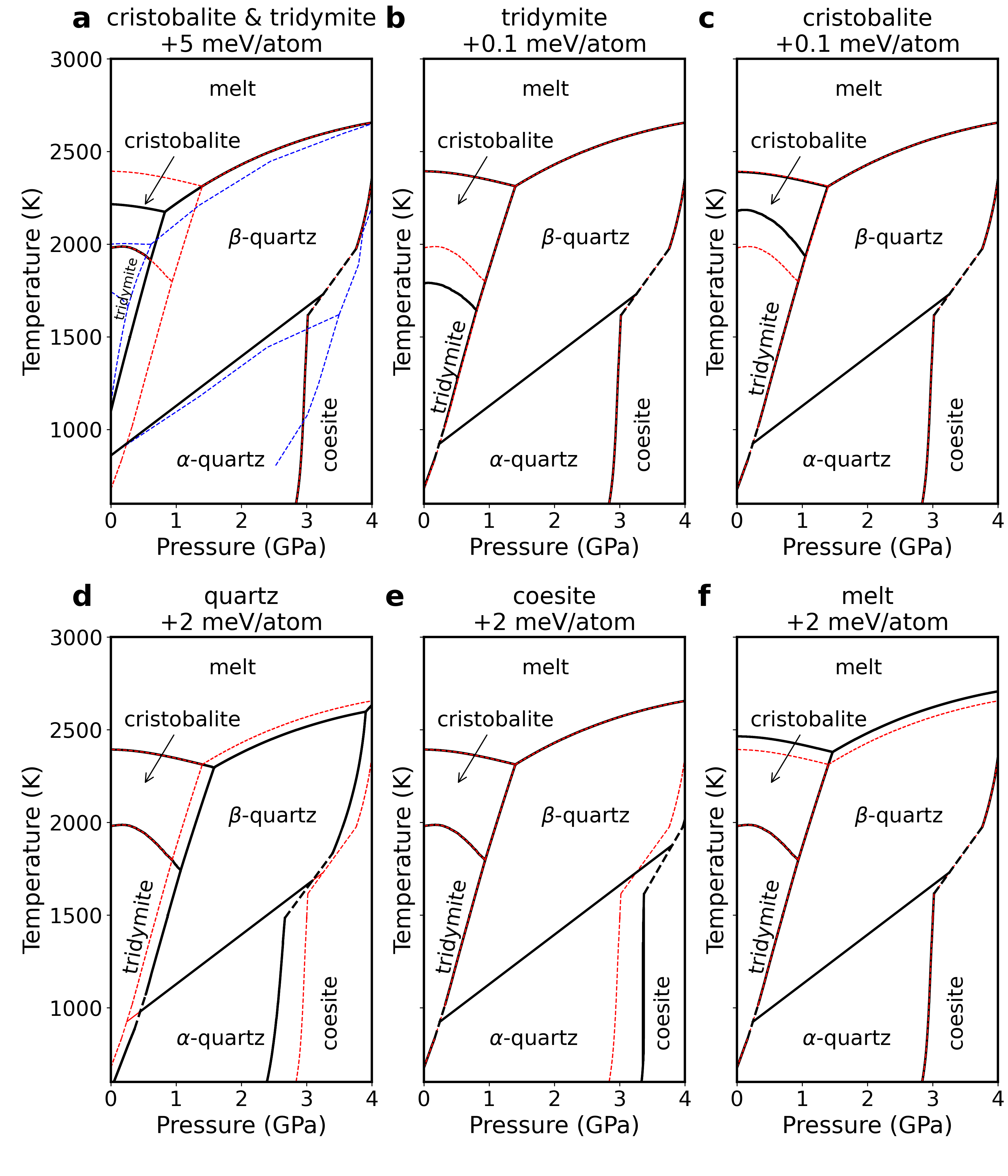}
    \caption{\textbf{Uncertainty of the phase diagram.} Modified phase diagrams by making certain phases less stable. The red lines in the background correspond to the native phase diagram. (a) shows the phase diagram for the case that cristobalite and tridymite are 5 meV/atom less favourable. The blue lines correspond to the CALPHAD phase diagram.\cite{swamyThermodynamicAssessmentSilica1994} (b)~and (c)~show the phase diagram, when tridymite or cristobalite would have a 0.1 meV/atom higher energy. Similarly, (d)-(f) show the phase diagram if quartz, coesite or the melt would have a 2~meV/atom higher energy. }
    \label{fig:phasediagrams}
\end{figure*}

\begin{figure*}
    \includegraphics[width=17cm]{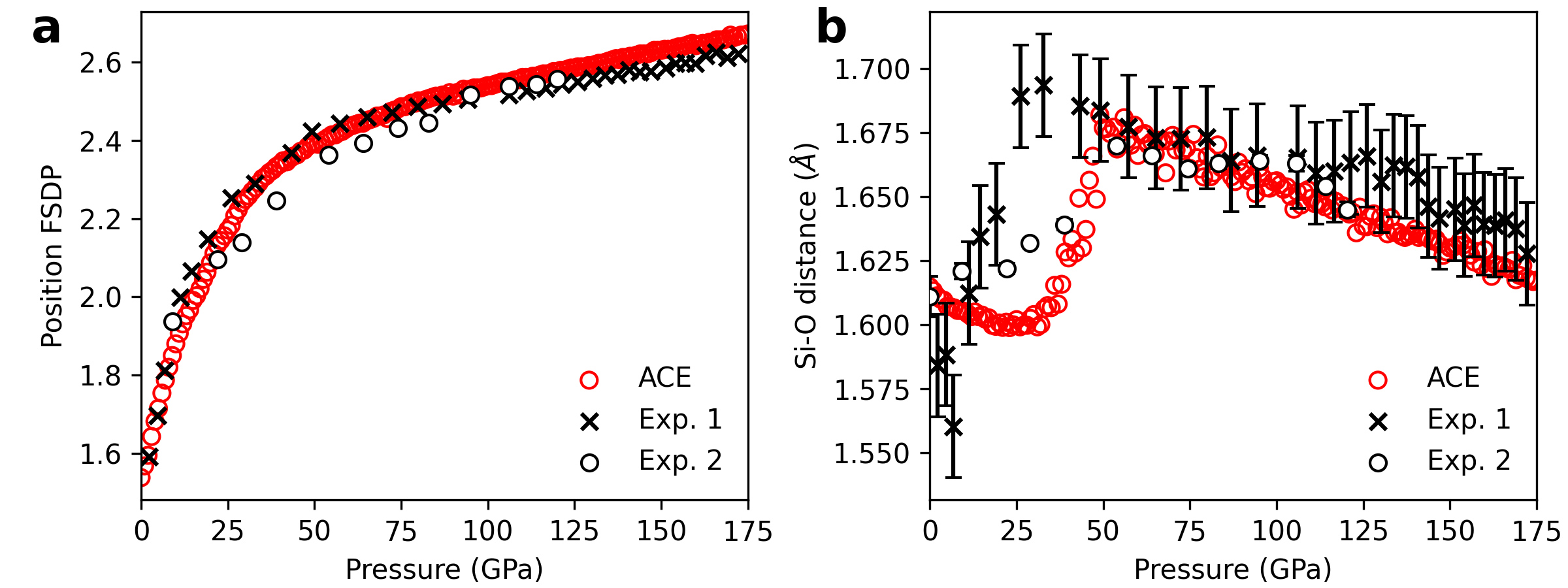}
    \caption{\textbf{Structural fingerprints of amorphous silica under high-pressure.} (a) Shows the position of the first sharp diffraction peak (FSDP) of the structure factor for various pressures. Similarly (b) shows the peak position of the first peak of the Si-O radial distribution function, which corresponds to the average Si-O bond distance. The experimental values are from Ref. \citenum{prescherSixfoldCoordinatedSi2017} and Ref. \citenum{konoStructuralEvolutionMathrmSiO2020}. }
\end{figure*}

\begin{figure*}
    \centering
    \includegraphics[width=7.5cm]{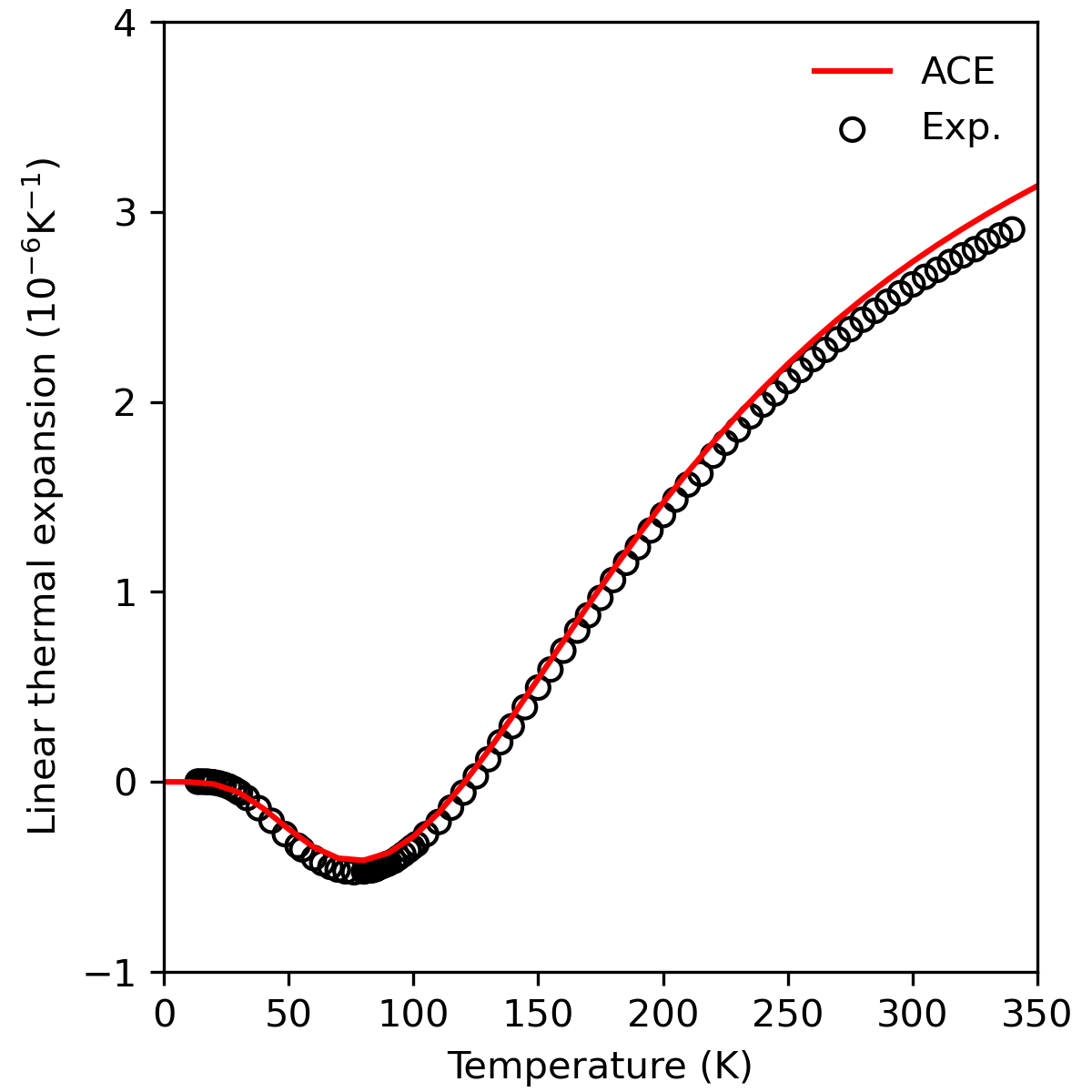}
    \caption{\textbf{Linear thermal expansion coefficient of diamond} calculated with ACE using the quasi-harmonic approximation \cite{togoFirstprinciplesPhononCalculations2010} as implemented in phonopy \cite{togoFirstPrinciplesPhonon2015,togoFirstprinciplesPhononCalculations2023} compared to experimental values from Ref. \citenum{lyonLinearThermalExpansion1977}.}
    \label{fig:thermalexpansion}
\end{figure*}

\begin{figure*}
    \centering
    \includegraphics[width=14cm]{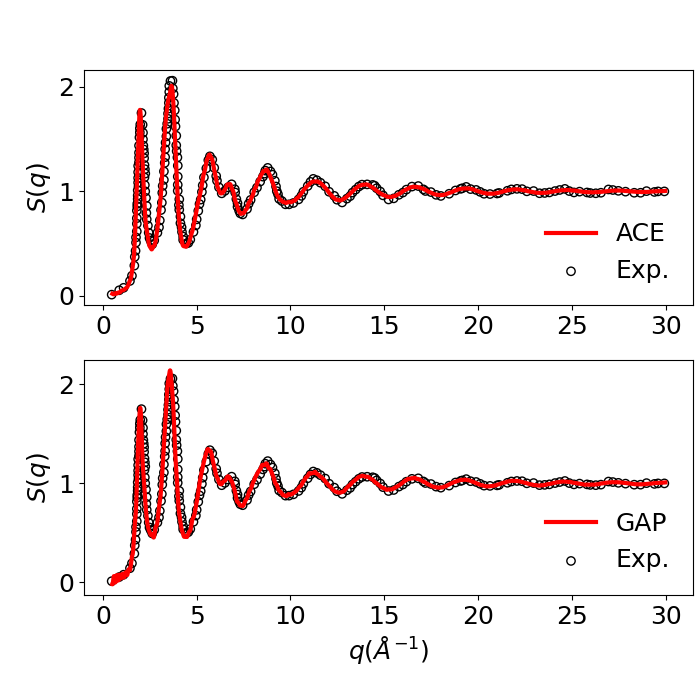}
    \caption{\textbf{Structure factor of amorphous silicon.} The structure factor of an amorphous silicon structure generated with the ACE potential compared to the experimentally determined structure factor from Ref. \citenum{laaziriHighenergyXrayDiffraction1999} and the structure factor of an 4000 atoms amorphous silicon structure generated by the GAP for silicon (lower panel). \cite{deringerRealisticAtomisticStructure2018a} The latter structure was equilibrated at 300~K with the GAP to determined the structure factor.}
    \label{fig:structurefactor}
\end{figure*}

\begin{figure*}
    \centering
    \includegraphics[width=7.5cm]{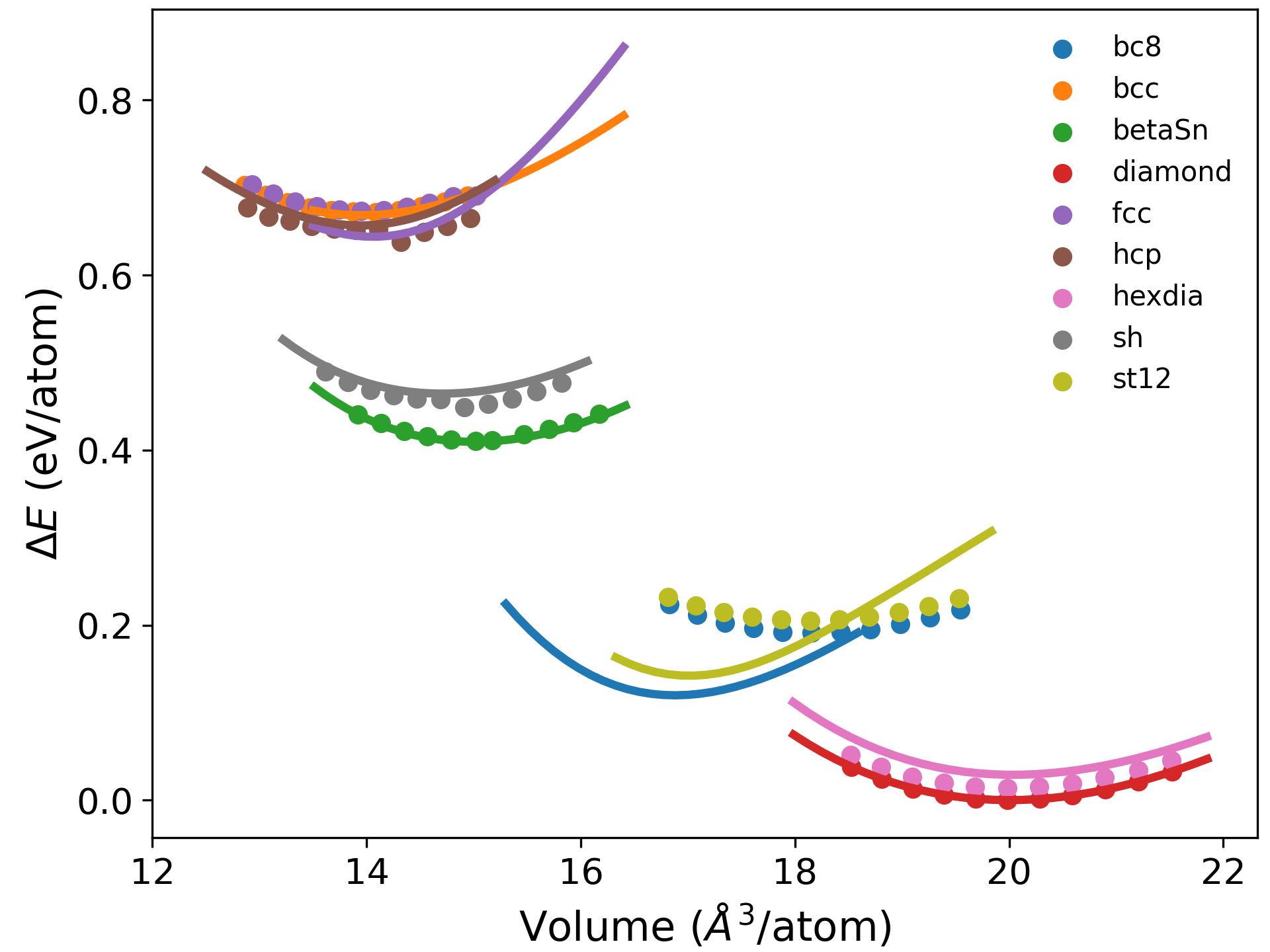}
    \caption{\textbf{Hydrostatic energy volume curves of various silicon polymorphs.} The dots indicate the relaxed DFT energies for each polymorph, while the lines are the energy volume curves of the ACE.}
\end{figure*}

\begin{figure*}
    \centering
    \includegraphics[width=10cm]{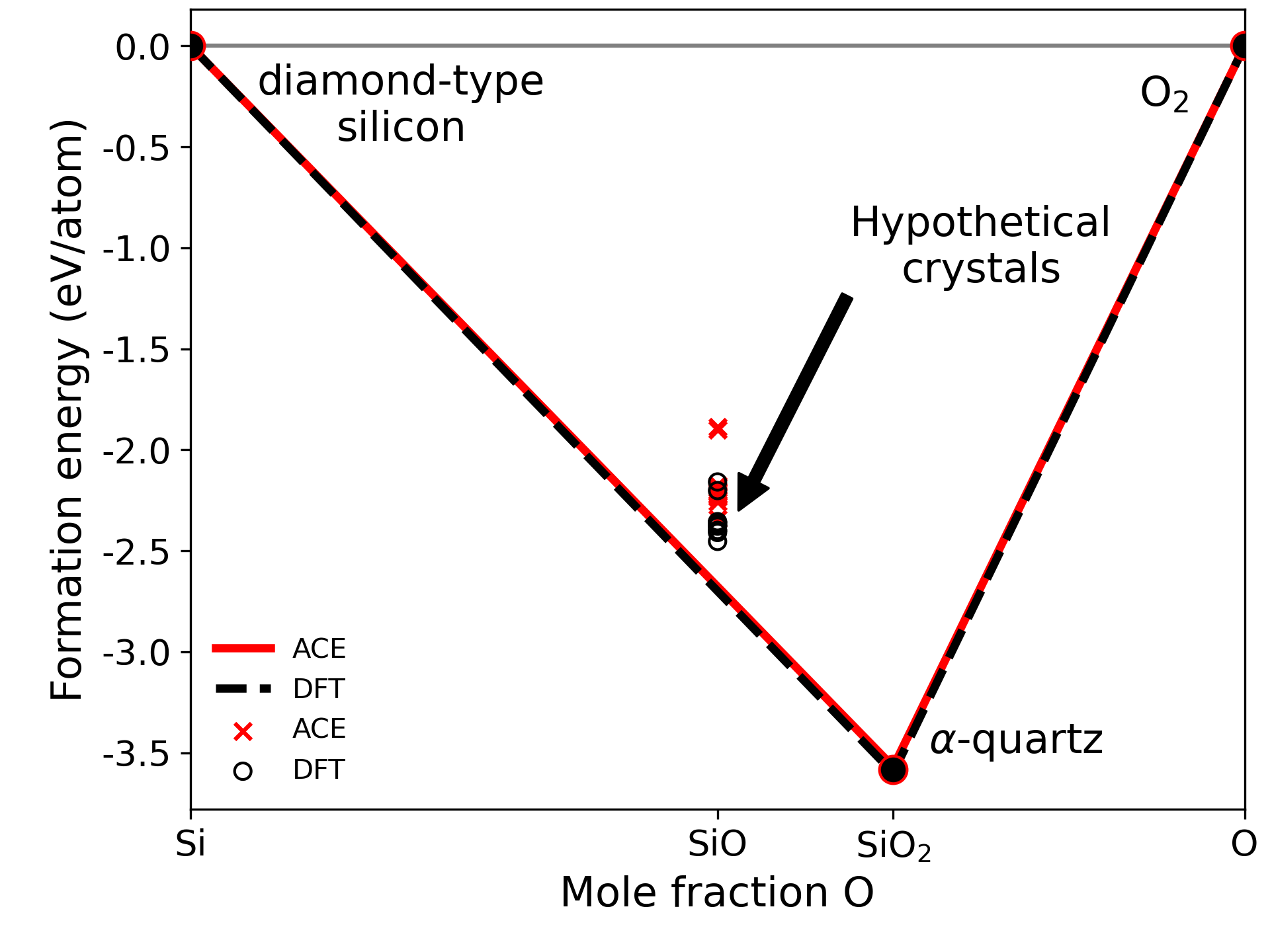}
    \caption{\textbf{Convex hull of the Si-O system.} For the stochiometric composition SiO, we added the formation energies of the hypothetical ambient pressure crystal structures from Ref. \citenum{alkaabiSiliconMonoxideAtm2014}. The energies of these structures predicted by the ACE are marked in red, while the DFT results are marked in black. None of these structures have been included in the training database.}
    \label{fig:econvex_hull}
\end{figure*}

\begin{figure*}
    \includegraphics[width=17cm]{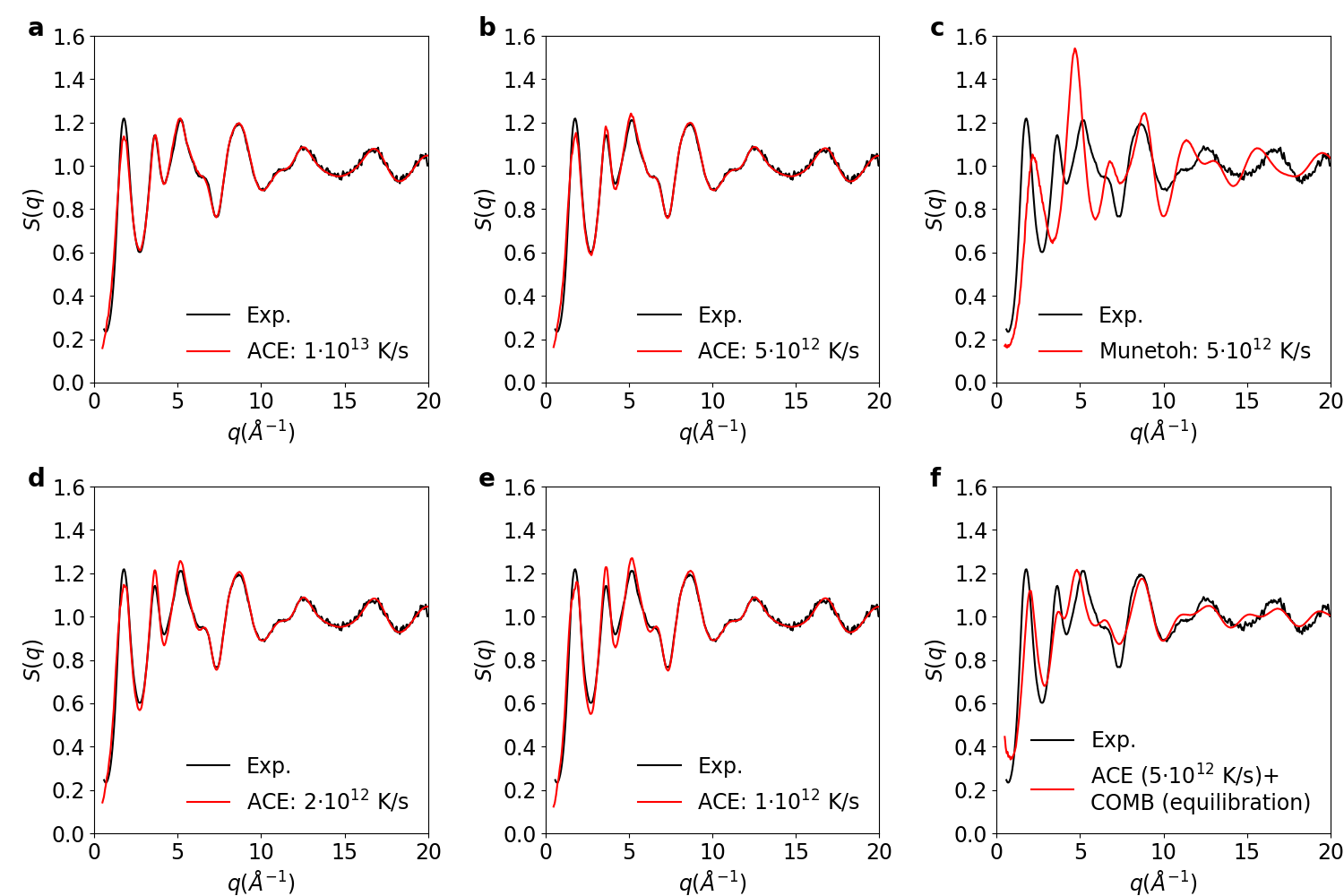}
    \caption{\textbf{Structure factors of SiO.} (a-b,d-e) are showing structure factors of SiO model structures generated by different quench rates using the ACE potential. The corresponding structure pictures are shown in the main part. (c) is showing the structure factor of a structure, which was generated with the same protocoll, however, with the Munetoh potential.\cite{munetohInteratomicPotentialSi2007b} (f) is showing the structure factor of the structure from (b), however, after relaxation with the COMB potential for the SiO system.\cite{shanSecondgenerationChargeoptimizedManybody2010}}
\end{figure*}

\clearpage

\section{Database}

\begin{table*}[]
    \centering
    \caption{\textbf{Composition of the total database} subdivided in various structure types with various compositions. We also show, which parts of the database are taken from references and which parts are new. For the structures, which are collected by active learning, we indicate which active learning type was used. We also give the fitting weights for each of the structure types.}
    \begin{tabular}{ccccccc}
        \hline
        \hline
        Configuration type & Composition & Reference & Active learning type & Structures & Atoms & Weights  \\
        \hline
        crystalline &SiO$_2$& SiO$_2$ GAP + this work &---& 2,620 & 281,820 & 100 \\
        amorphous &SiO$_2$& SiO$_2$ GAP &---& 313 & 60,096 & 1\\
        half-quenched &SiO$_2$& SiO$_2$ GAP &---& 311 & 59,712 & 1\\
        liquid &SiO$_2$& SiO$_2$ GAP &---& 313 & 60,096 & 1\\
        crystalline (main) &Si& Si GAP & --- & 1,257 & 38,680 & 100\\
        amorphous &Si& Si GAP & --- & 159 & 29,632 & 1\\
        liquid &Si& Si GAP & --- & 76 & 5,312 & 1\\
        surfaces &Si& Si GAP & --- & 214 & 22,066 & 1\\
        defects &Si& Si GAP & --- & 423 & 74,548 & 1\\
        various  (e.g. high energy crystal) &Si& Si GAP + this work & --- & 505 & 2,556 & 1\\
        quenched &SiO$_2$& this work & small-scale & 385 & 19,008 & 1\\
        quenched &SiO$_2$& this work & large-scale & 417 & 53,208 & 1\\
        vacancies &SiO$_2$& this work & --- & 278 & 56,520 & 1\\
        vacancies &SiO$_2$& this work & large-scale & 780 & 121,836 & 1\\
        high-pressure crystals &SiO$_2$ & this work & --- & 400 & 19,080  & 1 \\
        high-pressure amorphous & SiO$_2$ & this work & small-scale & 166 & 31,872 & 1 \\
        high-pressure amorphous & SiO$_2$ & this work & large-scale & 407 & 120,246 & 1\\
        surfaces & SiO$_2$ & this work & --- & 603 & 48,477 & 1\\
        surfaces & SiO$_2$ & this work & small-scale & 28 & 1872 & 1\\
        surfaces & SiO$_2$ & this work & large-scale & 167 & 8466 & 1\\
        crystalline-amorphous interfaces & Si+SiO$_2$ & this work & --- & 457 & 31036 & 1\\
        quenched & Si+SiO$_2$ & this work & small-scale & 457 & 31036 & 1\\
        quenched & Si+SiO$_2$ & this work & large-scale & 430 & 71821 & 1\\
        clusters (dimers, larger SiO clusters) & various & SiO$_2$ GAP + this work & ---/large-scale & 611 & 24,900 & 1 \\ 
        \hline
        \hline
    \end{tabular}
    \label{tab:database}
\end{table*}

The composition of the database (DB) and the weights used for the fit are
summarised in Supplementary Tab. \ref{tab:database}.
The corresponding energy and force range can be seen in the scatter plot in Supplementary Figure \ref{fig:scatter_plot}. 
The DB contains a large fraction of structures used for training the SiO$_2$ GAP-22 \cite{erhardMachinelearnedInteratomicPotential2022} (27\%) as well as the Si GAP-18 \cite{bartokMachineLearningGeneralPurpose2018} (22\%).
Structures added in this work represent high-pressure SiO$_2$ phases (both crystalline and amorphous), surfaces and charge-neutral vacancies,
as well as Si/SiO$_2$ interfaces, quenched structures of Si-SiO$_2$ mixtures and clusters of [SiO]$_x$.
These additional structures were partially designed on purpose and partially  obtained by active learning (AL),
either using small-scale models and the AL technique implemented in the MLIP package \cite{novikovMLIPPackageMoment2020}
or using commitee voting coupled to our amorphous matrix embedding, see Supplementary Sec. \ref{sec:ame}.
In both techniques we sample configuration space by molecular dynamics simulatons (MD).

\begin{figure*}
    \centering
    \includegraphics[width=17cm]{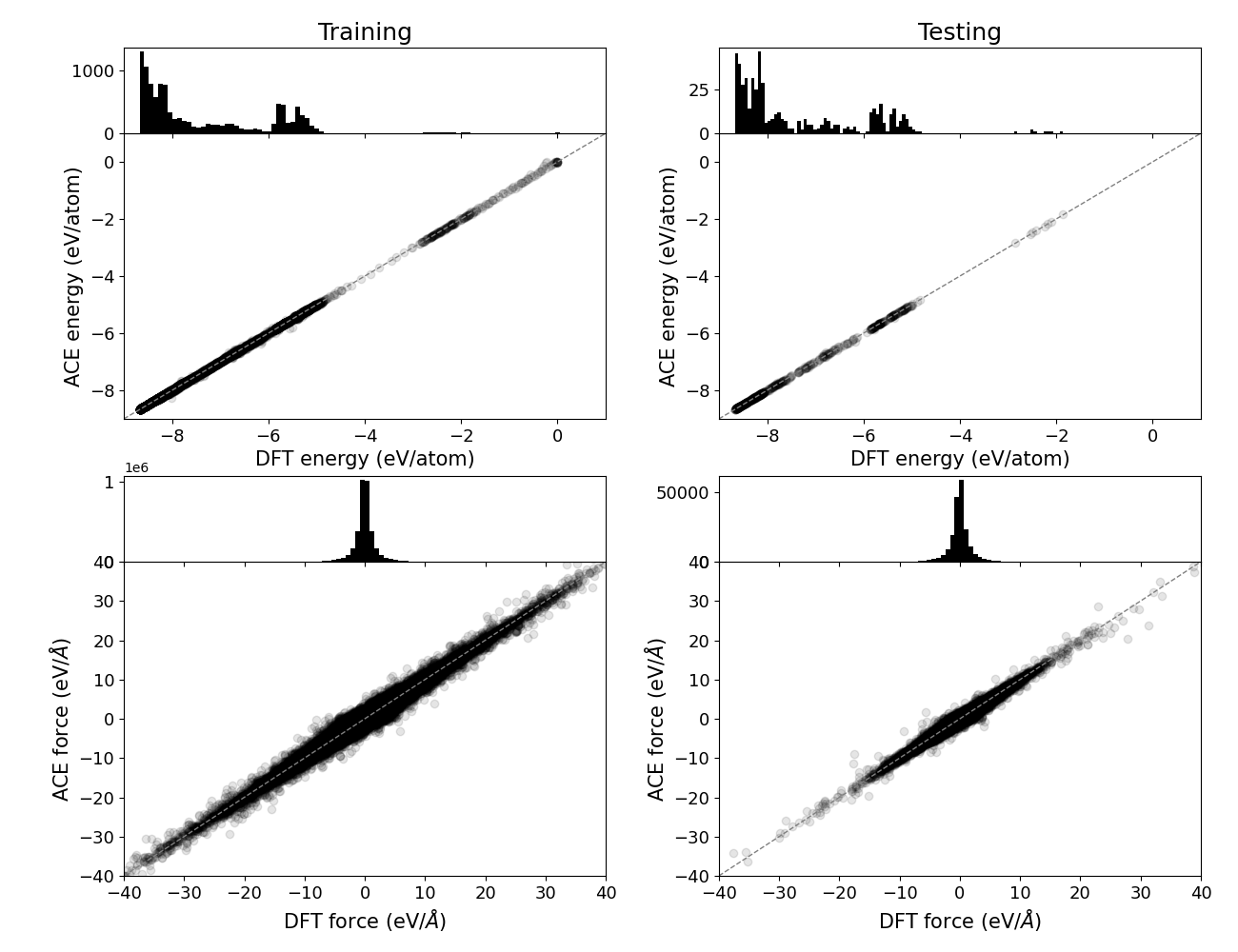}
    \caption{\textbf{Scatter plot} for the forces (bottom) and energies (top) of the training and testing set. The total dataset was randomly divided into the training (95\%) and the testing set (5\%). On top of the scatter plots we show the distribution of data points.}
    \label{fig:scatter_plot}
\end{figure*}

In the following details concerning the structure generation and related simulation parameters will be given.

\subsection{Manually prepared data}
Most of the manually prepared data is based on crystal structures or other input structures, which have been rattled and deformed.
If not stated differently, these structures have been prepare by applying variations of +/-~2.5~\% to the lattice parameters a,b and c and additionally by applying variations of +/-~5~\% to the angles $\alpha$, $\beta$ and $\gamma$. 
Afterwards the atoms have been displaced using the \texttt{ASE} \textit{rattle} function with a standard deviation of 0.01 \AA.\cite{larsenAtomicSimulationEnvironment2017}

\subsubsection{Crystalline  silica structures}
We added rattled and deformed crystalline supercells of the polymorphs $\alpha$-quartz, $\beta$-quartz,  moganite, coesite, stishovite, chabazite, $\alpha$-cristobalite, $\beta$-cristobalite,  low temperature tridymite, $\beta$-tridymite and tridymite in the C222$_1$ and P2$_1$2$_1$2$_1$ modifications. 

\subsubsection{fcc and hcp silicon}
We added fcc and hcp structure of silicon under high-compression to the database to improve the repulsive behaviour of the potential at short distances. 

\subsubsection{High pressure silica}
We included simple unit cells, as well as, 2x2x2 supercells of pyrite and $\alpha$-PbO$_2$-type silica in our database. 

\subsubsection{Silica surfaces}
Table \ref{tab:surfaces} shows a list of surfaces, which we included into our database. 
Each of these surfaces has been included 30 times in the database, 10 times by rattling the structures with an average displacement of 0.01~\AA, 10 time with an average displacement of 0.05~\AA~ and 10 times with 0.1~\AA. 

\begin{table} [H]
    \centering
    \caption{\textbf{Crystalline structures with surface orientations}, which are included in the database.}
    \begin{tabular}{cc}
        \hline
        \hline
        Polymorph & Surfaces \\
        \hline
         $\alpha$-quartz & (001), (110), (100), (210) \\
         $\alpha$-cristobalite & (001), (100), (110) \\
         low temperature tridymite & (100), (001) \\
         $\beta$-tridymite & (001), (110), (100), (210) \\
         $\beta$-cristobalite & (100), (110), (111) \\
         moganite & (100), (010), (001) \\
         \hline
         \hline
    \end{tabular}
    \label{tab:surfaces}
\end{table}

Additionally, we modified the surface terminations of several of these structures and rattled them 10 times by 0.01~\AA. 
Amorphous surface structure models have been created by using amorphous bulk samples and cutting them to create an artificial open surface. 

\subsubsection{Silica vacancies}
Vacancy structures are created by removing randomly atoms according to one formula unit of SiO$_2$ from $\alpha$-quartz and amorphous structure models.

\subsubsection{Crystalline-amorphous Si-SiO$_2$ interfaces}
Crystalline-amorphous interfaces between silicon and silica have been created by merging two crystalline bulk supercells together. 
One of these supercells was kept fixed, while the other was melted.
A repulsive wall was used to prevent atoms from the molten phase to diffuse into the crystalline phase.

\subsubsection{Clusters}
[SiO]$x$ clusters have been added in this work. 
These have been generated by iteratively performing MD simulations-. 
We started by an [SiO]$_2$ cluster and added five snapshots from five MD simulations (so in total 25) to the database.
Within these MD simulations the temperature was increased from 100 to 2000~K. 
After adding the structures to the database, we refitted the potential and repeated the simulations with clusters containing one additional formula unit of SiO resulting in a [SiO]$_{i+1}$ cluster. 
We repeated the process up to a system size of [SiO]$_{32}$. 

\subsection{Small-scale active learning}

In the small-scale active learning we used tools implemented in the \texttt{MLIP} package.\cite{novikovMLIPPackageMoment2020}
We performed MD simulations and recognised new structure using the maxvol algorithm as implemented in that code. 
We used an extrapolation threshold of 1.5 and a stopping threshold of 3.0.
This active learning process was repeated iteratively. 
We performed around 100 simulations in parallel, collected newly found structures and reduced them to the most relevant by the maxvol algorithm. 
Then we calculated corresponding forces and energies by DFT, refitted the potential and repeated the MD simulations.

In the following we will briefly give the simulation details of the performed MD simulations. 
We performed quenching simulations under various conditions. 
All quenching simulations have in common, that they start with a randomization phase for 10 to 100~ps at 6000~K and under NVT conditions. 
This is followed by an equilibration at 4000~K under NPT conditions for 100~ps and subsequent quenching to room temperature with various rates (10$^{12}$ to 10$^{14}$ K/s).

\subsubsection{Quenching under ambient pressure (SiO$_2$ and SiO$_x$)}

We performed quenches with the SiO$_2$ compositions and mixed SiO$_x$ compositions under ambient pressure using the quenching procedure explained above (external pressure of 0~GPa). 
For the SiO$_2$ simulations we used various crystalline unit- and supercells as input. 
For the mixed simulations we stacked together various SiO$_2$ and Si crystalline cells to achieve different compositions. 

\subsubsection{High-pressure amorphous silica}

Similar to the ambient pressure simulations, we also performed quenching simulations for the high-pressure phases. 
However, this time we used pressure between 0-200~GPa in the NPT part. 
After convergences of the quenching simulations we performed compression simulations using amorphous input structures under NPT conditions starting from a pressure of 0~GPa going to a pressure of 200~GPa. 
The temperature was held constant at a random value between 0 and 1000~K. 

\subsubsection{SiO$_2$ surfaces}

We used the surface structures, which have been created manually, as input for MD simulations.
In the MD simulations we annealed the surfaces under NVT conditions from 50~K to 3000~K and cooled them back down to 50~K afterwards.

\subsection{Large-scale active learning}

In the large-scale MD simulations we used a committee-error to examine the uncertainty of each atom. 
Based on the committee-error we decide whether a structure is needed to be added to the database. 
In the case that we want to add the structure we extract a small-scale model by amorphous matrix embedding. 
In this section we will briefly explain the amorphous matrix embedding method and then give the details for the MD simulations.

\subsubsection{Amorphous matrix embedding}
\label{sec:ame}

Amorphous matrix embedding is used to enable active learning (AL) on large-scale systems.
For DFT calculations, systems sizes are essentially restricted to a few hundreds of atoms.
Therefore conventional AL is performed using models of this size.
During large-scale simulations, however, local atomic environments may partially deviate from 
environments encountered in small-scale simulations.
It is therefore desirable to include environments that are likely to appear only in large-scale simulations.
Amorphous matrix embedding is a technique that allows us to do so.

The basic idea of the amorphous matrix embedding technique is to locally identify unknown environments in a large-scale simulation,
extract this environment including its surrounding (a total of 200 - 400 atoms),
freeze the atoms in the relevant environment and amorphize the surrounding using periodic boundary conditions,
see Fig. 1b in the research paper.
As measure for the uncertainty of an environment, we use committee voting of several moment-tensor potentials (MTPs)
with respect to forces on a particular atoms.
In particular, we define the uncertainty $u_{\alpha}$
\begin{equation}
    u_{\alpha} = \sqrt{\sum_{i=x,y,z}\text{STD}(F_{\alpha,i}^{(1..N)})^2}.
\end{equation}
Here $F_{\alpha,i}^{(1..N)}$ is the $i$-th component of the force acting on atom $\alpha$ as obtained by one of the committee members (potentials)
for particular configuration along a trajectory; STD denotes the standard deviation.

Based on this, the amorphous matrix embedding is performed now according to the following procedure.
First, we loop over all atoms and find atoms with an uncertainty above a certain threshold (1-2~eV/\AA).
In principle, this indicates that the local environment of the atom is unknown to the potential.
However, if the uncertainty is very high, the structure might be very unfavourable and contain atoms with very high forces.
This would worsen the DFT convergence and afterwards the potential fit. 
Therefore, we introduce an upper threshold (5~eV/\AA), which is not limited to the atom itself, but also to atoms in the direct environment. 
In the case that one of the atoms exceeds that threshold, we do not use this environment. 
Another issue is that similar structures might be picked in subsequent steps of the MD trajectory. 
We use smooth overlap of atomic positions similarities as implemented in \texttt{DScribe}\cite{himanenDScribeLibraryDescriptors2020} to add only structures, which are not similar to previously selected atoms (similarity$<$0.9-0.95).

Finally, for atomic environments that fullfill all the requirements, we create a new box with a box size of 13~\AA~plus an additional margin of 1~\AA.
The reason for this box size is that it is larger than twice the cutoff while still being DFT feasible.
We cut out a cube with a side length of 13~\AA~around the atom of interest and fill this cube into the newly created box.
Depending on the scenario, the composition of the box is adjusted by deleting atoms outside of the region of interest.
To fix the artificial boundaries we annealed the atoms outside of the cutoff up to 2000~K to 6000~K while keeping the inner atoms fixed. 
Afterwards, the structures were quenched to 300~K. 
To prevent atoms from outside the cutoff to enter the cutoff area, we applied a repulsive potential centred around the atom of interest.

For surfaces further special adjustments had to be made.
If the atom was close to the surface, we estimated the normal vector of the surface by,

\begin{equation}
    \textbf{n}_i \approx \sum_j (\textbf{x}_i-\textbf{x}_j) \text{  if distance}(i,j) < \text{cutoff}.
\end{equation}

Here $\textbf{x}_j$ is the position of atom $j$. 
The box, which needed to be extracted, was then rotated in a way that the normal vector was oriented along the z-direction.
Additional vacuum layer of 5~\AA~along each direction have been placed into the direction of the normal vector. 
After the molecular dynamics simulations some of the surfaces structures have not been periodically connected into the x and y directions.
These cases were handled instead of slab structures as cluster structures and additional vacuum into the x- and y-direction was added, to be able to use the dipole correction in VASP.

\subsubsection{Quenching under ambient pressure (SiO$_2$ and SiO$_x$)}

We used the same protocol for quenches as mentioned in the small-scale active learning part. 
As input structure for the SiO$_2$ simulations we used $\beta$-cristobalite supercells, while for the SiO$_x$ simulations we used supercells of the cell used for the small-scale simulations.
For the latter one an exemplary structure is shown in Supplementary Fig. \ref{fig:input_siO}.

\begin{figure}[H]
    \centering
    \includegraphics[width=6cm]{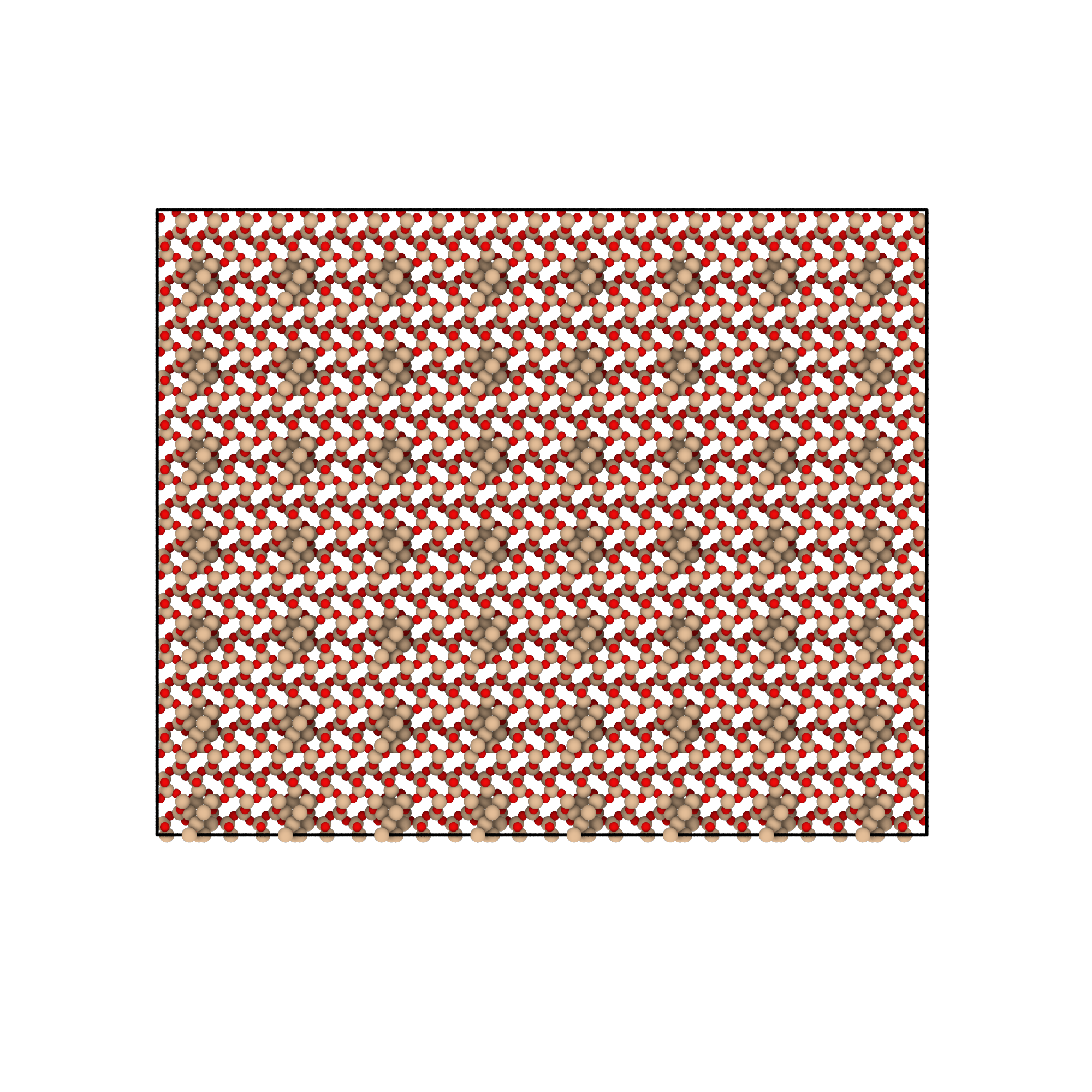}
    \caption{\textbf{Initial structure for the large-scale SiO$_x$ interpolation models.} The structure is a supercell of $\alpha$-cristobalite structures merged with silicon structures of diamond-type.}
    \label{fig:input_siO}
\end{figure}

\subsubsection{Silica surfaces}

To perform the large-scale active learning, we used the same standard quenching protocol as mentioned above. 
However, the NPT part was replaced by a part, where the initially liquid structure was constantly strained to larger volumes inducing pores into the structure. 

\subsubsection{Vacancies}

As input structures we used quartz supercells and amorphous structures containing around 65,000 atoms including 50 formula units of randomly selected vacancies. 
These structures have been annealed from room temperature to 3000~K and back to room temperature. 

\subsubsection{Compression to very--high--pressures}

We used amorphous models containing 65,000 atoms, which we compressed in MD simulations up to 200~GPa at a constant randomly selected temperature between 0 and 1000~K.

\subsubsection{Clusters}

We used a box containing 10,000 SiO molecules in gaseous state with a density of 0.011~g/cm$^3$ as input structure. 
This box was compressed to a density of $\approx$2~g/cm$^3$, while keeping the temperature constant at 1400~K.

\section{Calculation of physical properties}

\subsection{Phase diagram calculation}

For the determination of the phase diagram we used calphy \cite{menonAutomatedFreeenergyCalculation2021}, which implements thermodynamic integration using reversible scaling \cite{dekoningOptimizedFreeEnergyEvaluation1999} and nonequilibrium calculations of free energy differences. 
The basic idea is that we are switching the Hamiltonian of our system continuously to the Hamiltonian of a references system with known free energy. 
Calphy uses either the Einstein crystal for crystalline structures or the Uhlenbeck-Ford model \cite{paulaleiteUhlenbeckFordModelExact2016} for liquid systems.
Finally, we receive an absolute Gibbs free energy for a certain temperature and pressure.

This Gibbs free energy is used as a reference energy, from which we start to perform thermodynamic integration using reversible scaling. 
By this we are able to find the temperature dependence of the Gibbs free energy for a given pressure or although not used here also the pressure dependence for a given temperature.
Details of the theoretical background can be found in the corresponding publication.\cite{menonAutomatedFreeenergyCalculation2021}

We used input structures with a cell size of around $\approx$ 15,000 atoms.
For the calculations we used 50,000 equilibration steps and 800,000 switching steps for switching between our systems and the corresponding reference systems. 
For the thermodynamic integration we used 300 steps/K. 
The force constants for the Einstein crystal were set to 2~eV/\AA$^2$ for oxygen and 4~eV/\AA$^2$ for silicon, to avoid numerical instabilities. 
We carefully checked the dependence of the final results on the force constants and can exclude notable impact ($\Delta G<$0.01 meV/atom). 

\begin{figure*}
    \centering
    \includegraphics[width=17cm]{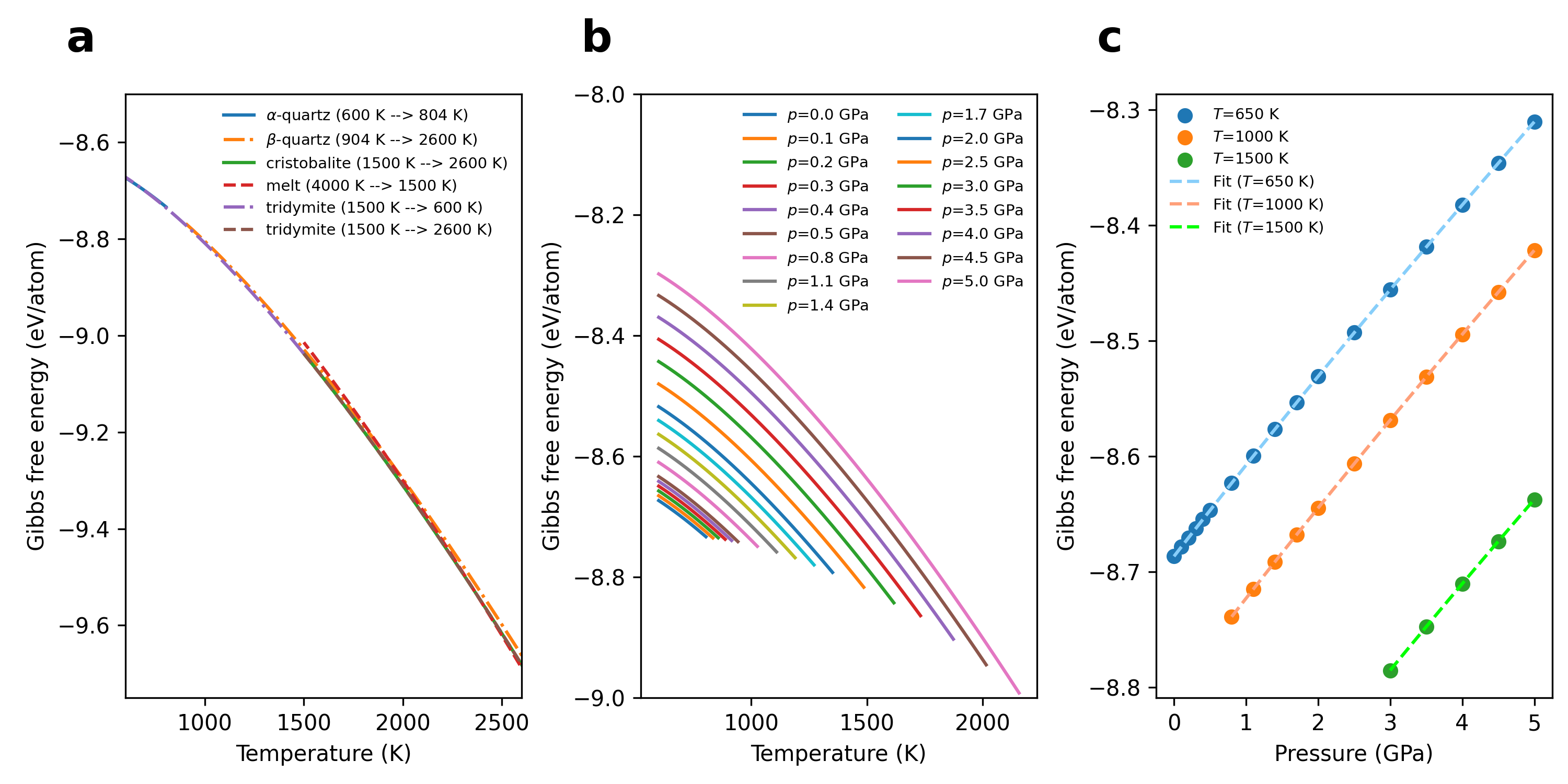}
    \caption{\textbf{Free energies by thermodynamic integration.} (a) Shows the Gibbs free energies calculated by \texttt{calphy} for various silica polymorphs for different temperature ranges at 0~GPa pressure. (b) shows the Gibbs free energy results for $\alpha$-quartz at various pressures. (c) shows a polynomial fit of the Gibbs free energy in dependence of the pressure for $\alpha$-quartz at different temperatures.}
    \label{fig:thermodynamicintegration}
\end{figure*}

In our work, we sampled the free energies in a temperature range between 600~K and $\sim$2600-3000~K for various polymorphs. 
An example is shown in Supplementary Fig. \ref{fig:thermodynamicintegration}a.
Here, we show the Gibbs free energies for four different structure types of silica at a pressure of 0~GPa.
For cristobalite we started integration at 1500~K up to shortly above the melting point.
The tridymite interval was split into two integrations intervals to achieve higher accuracy.
One integration was performed from 1500~K to 600~K and the other integration was performed from 1500~K to 2600~K. 
The melt was integrated down from 4000~K to 1500~K. 
Unfortunately, we cannot integrate quartz in the same way since $\alpha$-quartz is dynamically switching to $\beta$-quartz in the simulations.
Therefore, the $\alpha$-quartz/$\beta$-quartz transition was determined using classical molecular dynamics simulations. 
This is illustrated in Supplementary Fig. \ref{fig:abquartztrans}a-b. 
$\alpha$-quartz as well as $\beta$-quartz input structures have been equilibrated in molecular dynamics simulations at various temperatures for 100~ps. 
Already around 30~K away from the transition points the final densities agree very well. 
The transition temperature was determined by finding the maximum slope of the following approximating function,

\begin{figure*}
    \centering
    \includegraphics[width=17cm]{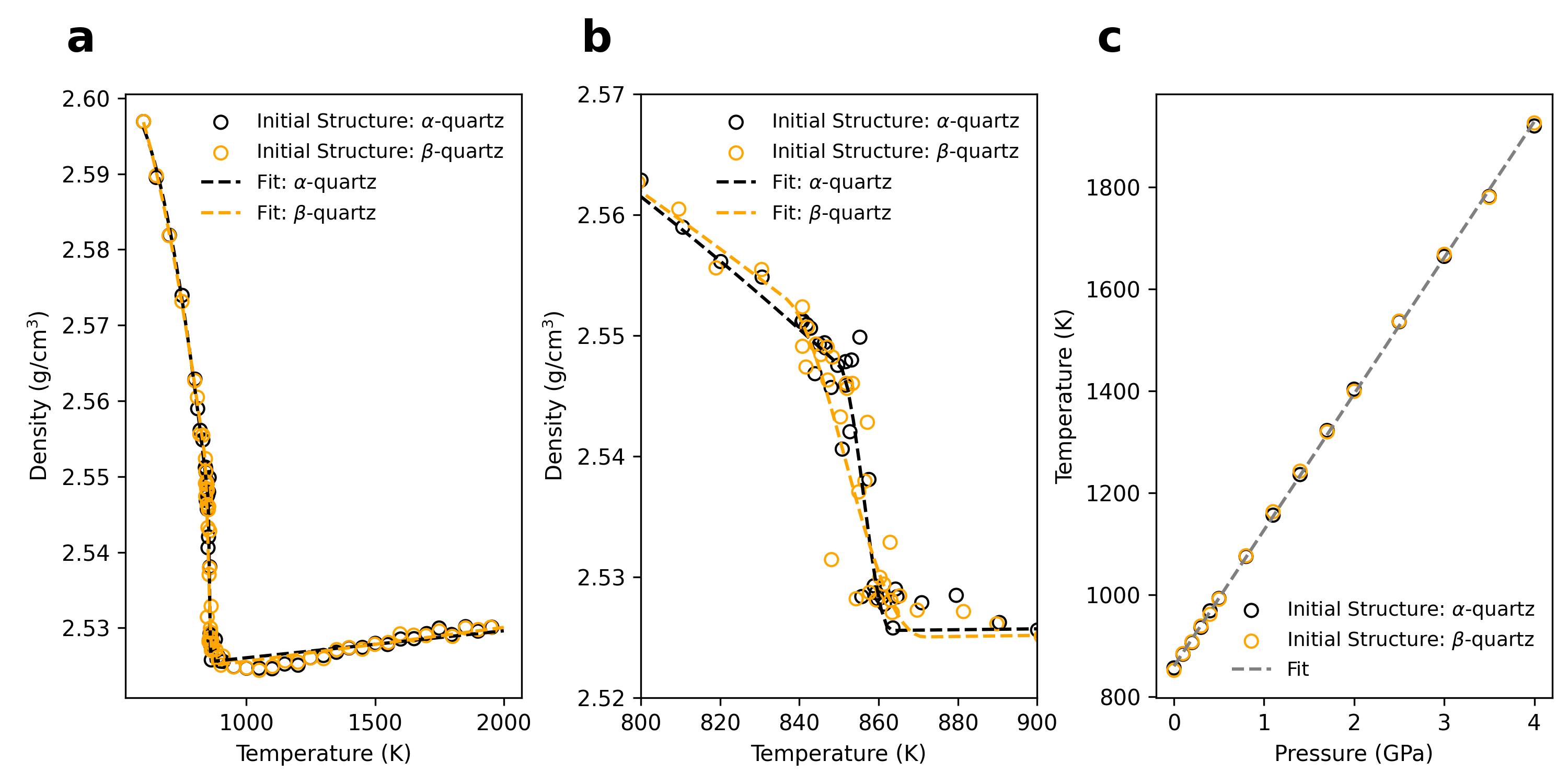}
    \caption{\textbf{Determination of the $\alpha$/$\beta$-quartz transition} (a) shows the density of quartz depending on the temperature for a pressure of 0 GPa. It can be seen that there is a distinct transition from $\alpha$-quartz to $\beta$-quartz. (b) shows the data of (a) limited to a temperature range of 800-900~K. It can be seen that the transition from $\alpha$- to $\beta$-quartz appears around 850~K. Moreover, we see the good agreement of the fitting functions with the data. In (c) we show the transition temperature from $\alpha$-quartz to $\beta$-quartz in dependence of the pressure.}
    \label{fig:abquartztrans}
\end{figure*}

\begin{equation}
    \rho (T) \approx \begin{cases}
    a\cdot x^2+b\cdot x+c & x < T_1 \\
    0.5\cdot \cos{\left( \dfrac{x-T_1}{T_2-T_1}\pi\right) }+0.5 & T_1 < x < T_2 \\
    d\cdot x+e & x > T_2 \\
    \end{cases}.
\end{equation}

The parameters $a$, $b$, $c$, $d$, $e$, $T_1$ and $T_2$ are fitted to the results from the molecular dynamics simulations.
This was repeated for several pressures.
Corresponding transition temperatures in dependence of the pressure are shown in Supplementary Fig. \ref{fig:abquartztrans}c. 
A line is shown, which was fitted to the average transition temperatures.
This line was also used in the phase diagram plots in Figure 2. 
Based on this transition line we performed thermodynamic integration for $\alpha$-quartz from 600~K to 50~K below the $\alpha$-$\beta$-quartz transition line, while for $\beta$-quartz we started 50~K above this line.
Therefore, we do not have data for quartz in a range of 100~K around the transition line.

This whole process was repeated for several pressures as it is indicated in Supplementary Fig. \ref{fig:thermodynamicintegration}b for $\alpha$-quartz. 
Corresponding to the pressure the temperature integration interval was adjusted for each polymorph.
Finally, to calculate the phase diagram, we fitted our data points to a polynomial of third order to achieve a equation for the Gibbs free energy dependence of the pressure.
This was done for temperatures between 600 and 2900~K in 1~K steps.
Exemplary, this is shown in Supplementary Fig. \ref{fig:thermodynamicintegration} for $\alpha$-quartz for temperatures of 650, 1000, 1500~K.

Supplementary Figure \ref{fig:phasediagrams} emphasises the difficulties of achieving accurate agreement between calculated and measured phase diagrams. We show how decreasing the stability of certain phases change the phase diagram compared to the original one. 
The original phase diagram is shown as thin red line in the background, while the modified one is shown by the black lines. 
For quartz, coesite and the melt we added 2 meV/atom to the free energy since the coexistence lines are less sensitive to a shift of the energy compared to cristobalite and tridymite. 
These are shown in Supplementary Fig. \ref{fig:phasediagrams}d-f.
For cristobalite and tridymite we added only 0.1 meV/atom.
The corresponding change of the transition lines is shown in Supplementary Fig. \ref{fig:phasediagrams}b-c. 
Additionally, in Supplementary Fig. \ref{fig:phasediagrams}a we show how the phase diagram changes, when we add 5 meV/atom to cristobalite and tridymite.
The additional blue line in the background shows the CALPHAD reference data from \cite{swamyThermodynamicAssessmentSilica1994}. 
There are several points we can conclude from this. 
First, we can conclude that there is a high uncertainty in the determination of the cristobalite-tridymite transition line. 
Especially, with respect to the error of potential to the corresponding DFT data ($\approx$1~meV/atom).
Second, it shows how essential the choice of the DFT exchange-correlation functional is, since depending on the functional this phase diagram would look totally different. 
The reason for this is that the difference between two exchange-correlation functionals can be more than 100 meV/atom for silica.\cite{erhardMachinelearnedInteratomicPotential2022}
Thirdly, we can see that also the SCAN exchange-correlation functional is not perfect. 
Destabilising cristobalite and tridymite would improve the match with the experimental phase diagram significantly.
Therefore, it seems like that the SCAN energies for these polymorphs are predicted slightly too low.

\subsection{Surface Energies}

Table II and Fig. 4 both in the main manuscript, show surfaces energies of silicon and silica. We consider only stoichiometric slabs. These surface energies $\gamma$ have been calculated by the following formula,

\begin{equation}
    \gamma = \dfrac{E_{\text{slab}}-N\cdot E_{\text{ref}}}{A},
\end{equation}

where $A$ is the total surface area, $N$ is the number of particles in the slab, $E_{\text{ref}}$ is the bulk reference energy and $E_{\text{slab}}$ is the potential energy of the slab. 
The reference energy of the $\alpha$-quartz surface is the energy of the minimised $\alpha$-quartz unit cell per atom, correspondingly the reference energy of the diamond surface is the minimised diamond unit cell. 
The reference energy of the amorphous sample is the bulk energy of the same relaxed amorphous structure without surfaces.

\subsection{Enthalpy}

Fig. 3 in the main manuscript shows the enthalpy for several phases. The enthalpy $H$ is given by,

\begin{equation}
    H(p) = E(V)+p(V) \cdot V,
\end{equation}

where $E$ is the inner energy, $p$ is the pressure and $V$ is the volume. 
The volume dependence of the energy has been determined by a Birch-Murnaghan fit to the energy-volume curve of each polymorph.
$p(V)$ was given by the corresponding derivative. 
The energy-volume curves have been calculated by variing the volume by $\pm$20\% for $\alpha$-quartz and coesite, by $\pm$25\% for stishovite and $\pm$30\% for all other phases. 
The corresponding structures have been structurally minimised allowing changes of the positions as well of the box shape, however, keeping the volume fixed.
Coordination numbers have been determined by the integral over the first peak of the partial Si-O radial distribution function.
The Si-O bond distances is given by the first peak position of the partial Si-O radial-distribution function.

\subsection{Thermal expansion coefficient}

Supplementary Figure \ref{fig:thermalexpansion} shows the thermal expansion coefficient of diamond. 
This was calculated within the quasi-harmonic approximation \cite{togoFirstprinciplesPhononCalculations2010} as implemented in \texttt{phonopy} \cite{togoFirstPrinciplesPhonon2015,togoFirstprinciplesPhononCalculations2023}. 

\subsection{Amorphous silicon}

Supplementary Figure \ref{fig:structurefactor} shows the structure factors of two amorphous silicon structures.
One generated by the GAP-18\cite{deringerRealisticAtomisticStructure2018a} and one generated by the ACE using a quench rate of 10$^{11}$~K/s with 8000 silicon atoms. 
The structure was randomized at 2500~K followed by a subsequent equilibration for 100~ps at 2000~K. 
Afterwards the structure was quenched to 500~K and then equilibrated at 300~K for 20~ps to determine the structure factor. 

\section{Performance of the ACE}

Supplementary Table \ref{tab:accuracy} shows a more comprehensive version of Tab. I in the main manuscript. 
The table shows that in nearly all cases the accuracy of the complex ACE potential is better or comparable to the accuracy of the GAP-22. 
Indeed, the forces are significantly more accurate, while the difference for the energy is not too large. 
Similarly, Supplementary Tab. \ref{tab:speed} shows that the ACE is at the same time more than 150 times faster than the GAP. 
Moreover, the speed is only slightly lower compared to the linear and Finnis-Sinclair like ACE.

\begin{table*}[]
    \caption{\textbf{Comparison of the accuracy} of the GAP\cite{erhardMachinelearnedInteratomicPotential2022} with various ACE models for the training and testing set, as well for a range of separate smaller datasets. The errors given in the table are root mean square errors for energies and forces. The unit of the energy error is meV/atom and of the force error meV/\AA.}
    \centering
    \begin{tabular}{lcccccccccccc}
    \hline
    \hline
         & & & & \multicolumn{2}{c}{\ce{SiO2}-GAP} & & \multicolumn{6}{c}{Si--O-ACE} \\
         & & & & \multicolumn{2}{c}{(Ref.~\citenum{erhardMachinelearnedInteratomicPotential2022})} & & \multicolumn{6}{c}{(This work)} \\
         \cline{5-6} \cline{8-13}
         & & & & & & & \multicolumn{2}{c}{Linear} & \multicolumn{2}{c}{F--S} & \multicolumn{2}{c}{Complex} \\
         & & & & & & & \multicolumn{2}{c}{$(N=1)$} & \multicolumn{2}{c}{$(N=2)$} & \multicolumn{2}{c}{$(N=8)$} \\
         & \textit{details} & $N_{\text{structures}}$ & $N_{\text{atoms}}$ & $\Delta E$ & $\Delta F$ & & $\Delta E$ & $\Delta F$ & $\Delta E$ & $\Delta F$ & $\Delta E$ & $\Delta F$ \\
    \hline
    Training & & 10853 & 1192984 & --- & --- & & 70.7  & 492  & 31.5 & 436 & 17.7 & 306 \\
    Testing  & & 571 & 65208 & --- & --- & & 51.5 & 490 & 33.1 & 431 & 16.7 & 305 \\
    \hline
    \ce{SiO2} crystals & $\alpha$-quartz,coesite,stishovite & 15 & 840 & 1.0 & 82 & & 0.8 & 74 & 1.1 & 62 & 0.9 & 45 \\
    \ce{SiO2} surfaces & amorphous & 5 & 540 & 14.9 & 178 & & 21.4 & 206 & 18.0 & 182 & 4.7 & 156 \\
    \hline
    a-\ce{SiO2} (ACE-MD)  & T=500,1500,3000~K & 15 & 2880 & 4.0 & 173 & & 8.0 & 277 & 7.4 & 260 & 3.2 & 176 \\
    a-\ce{SiO2} (CHIK-MD) & T=300~K, see Ref. \citenum{erhardMachinelearnedInteratomicPotential2022} &  5& 960 & 3.7 & 188 & & 4.1 & 270 & 5.1 & 267 & 2.2 & 192 \\
    a-\ce{SiO2} (GAP-MD)  & T=300~K, see Ref. \citenum{erhardMachinelearnedInteratomicPotential2022}&  5& 960 & 1.1 & 101 & & 10.3 & 132 & 9.8 & 120 & 4.6 & 96  \\
    a-\ce{SiO2} (BKS-MD)  & T=300~K, see Ref. \citenum{erhardMachinelearnedInteratomicPotential2022}&  5& 960 & 1.7 & 130 & & 3.3 & 210 & 3.2 & 185 & 1.3 & 132 \\
    a-\ce{SiO2} (Vashishta-MD) & T=300~K, see Ref. \citenum{erhardMachinelearnedInteratomicPotential2022}& 5 & 960 & 5.7 & 221 & & 9.9 & 330 & 10.0 & 325 & 3.1 & 224 \\
    a-\ce{SiO2} (Munetoh-MD) & T=300~K, see Ref. \citenum{erhardMachinelearnedInteratomicPotential2022}& 5 & 960 & 8.5 & 508 & & 75.9 & 591 & 55.3 & 552  & 22.0 & 371   \\
    \hline
    Amorphous Si  & T=500,1500,3000~K & 15 & 1920 & $>$ 1600 & $>$ 3200 & & 115.8 & 375 & 53.9 & 339 & 51.5 & 258  \\
    Mixed phases  & T=500,1500,3000~K & 15 & 1920 & $>$ 4200 & $>$ 3500 & & 37.8 & 710 & 35.0 & 635 & 38.0 & 431  \\
    High pressure & P=50,150~GPa& 10 & 1920 & 122.7 & 873 & & 15.1 & 476 & 5.6 & 359 & 4.6 & 236   \\
    \hline
    \hline
    \end{tabular}
    \label{tab:accuracy}
\end{table*}

\begin{table*}
    \caption{\textbf{Timing and speedup} of various ACE potentials compared to the GAP from Ref. \citenum{erhardMachinelearnedInteratomicPotential2022}. Timings have been obtained for 192 atoms cells over 100 time steps.}
    \centering
    \begin{tabular}{ccccccc}
    \hline
    \hline
                & GAP && \multicolumn{3}{c}{ACE}   \\
                & (Ref. \citenum{erhardMachinelearnedInteratomicPotential2022}) && \multicolumn{3}{c}{(This work)} \\
                \cline{2-2} \cline{4-6}
                &  && Linear & F-S & Complex \\
                &  &&(N=1) & (N=2) & (N=8) \\
    \hline           
         Timing ($\mu s$/(timestep $\cdot$ atom)) & 11037 & & 61 & 62 & 70 \\
         Speed--up to GAP   & 1 & & 181 & 178 &  158 \\
    \hline
    \hline
    \end{tabular}
    \label{tab:speed}
\end{table*}

\clearpage
\bibliographystyle{naturemag}
\bibliography{SiO_supplementary}